\documentclass[useAMS,usenatbib]{mn2e}
\usepackage{graphicx}
\usepackage{textcomp}
\usepackage{hyperref}
\usepackage{threeparttable}
\usepackage{longtable}
\usepackage[fleqn]{amsmath}
\usepackage{lscape}
\usepackage{pdflscape}
\usepackage{dcolumn}
\usepackage{rccol}
\usepackage{environ}
\usepackage{threeparttablex}
\usepackage{amssymb}
\newcommand{\mstellar}{\ensuremath{M_{\mathrm{stellar}}}}
\newcommand{\omatter}{\ensuremath{\Omega_{\mathrm{M}}}}

\newcommand{\col}{\ensuremath{\mathcal{C}}}
\newcommand{\absm}{\ensuremath{M_B}}
\newcommand{\scriptm}{\ensuremath{\mathcal{M}_B}}

\newcommand{\mbcorr}{\ensuremath{m_{B}^{\mathrm{corr}}}}
\newcommand{\mB}{\ensuremath{m_{B}}}

\newcommand{\mbmodel}{\ensuremath{m_{B}^{\mathrm{mod}}}}

\newcommand{\oii}{\ensuremath{\mathrm{O}\,\textsc{ii}}}
\newcommand{\oiii}{\ensuremath{\mathrm{O}\,\textsc{iii}}}
\newcommand{\nii}{\ensuremath{\mathrm{N}\,\textsc{ii}}}
\newcommand{\hii}{\ensuremath{\mathrm{H}\,\textsc{ii}}}
\newcommand{\Nifs}{$^{56}$Ni}
\newcommand{\nodata}{$\cdots$}

\defcitealias{2004MNRAS.348L..59P}{PP04}
\defcitealias{2008ApJ...681.1183K}{KE08}
\defcitealias{2002ApJS..142...35K}{KD02}

\title[The host galaxies of PTF SNe Ia]{The Host Galaxies of Type Ia
  Supernovae Discovered by the Palomar Transient Factory}
\author[Pan et al.]{Y.-C. Pan$^{1}$\thanks{E-mail:Yen-Chen.Pan@astro.ox.ac.uk}, M. Sullivan$^{2}$, K. Maguire${^1}$, I. M. Hook$^{1,3}$, P. E. Nugent$^{4,5}$,\newauthor D. A. Howell$^{6,7}$, I. Arcavi$^{8}$, J. Botyanszki$^{5,10}$, S. B. Cenko$^{9,19}$, J. DeRose$^{10}$,\newauthor H. K. Fakhouri$^{10,11}$, A. Gal-Yam$^{8}$, E. Hsiao$^{12}$, S. R. Kulkarni$^{13}$, R. R. Laher$^{14}$,\newauthor C. Lidman$^{15}$, J. Nordin$^{11,16}$, E. S. Walker$^{17}$, D. Xu$^{18}$\\
  $^{1}$Department of Physics (Astrophysics), University of Oxford, DWB, Keble Road, Oxford OX1 3RH, UK\\
  $^{2}$School of Physics and Astronomy, University of Southampton,Southampton, SO17 1BJ, UK\\
  $^{3}$INAF Osservatorio Astronomico di Roma, via Frascati, 33, 00040 Monte Porzio Catone, Roma, Italy\\
  $^{4}$Department of Astronomy, University of California, Berkeley, CA 94720-3411, USA\\
  $^{5}$Computational Cosmology Center, Lawrence Berkeley National Laboratory, 1 Cyclotron Road, Berkeley, CA 94720, USA\\
  $^{6}$Las Cumbres Observatory Global Telescope Network, Goleta, CA 93117, USA\\
  $^{7}$Department of Physics, University of California, Santa Barbara, CA 93106-9530, USA\\
  $^{8}$Department of Particle Physics and Astrophysics, Weizmann Institute of Science, Rehovot 76100, Israel\\
  $^{9}$Astrophysics Science Division, NASA Goddard Space Flight Center, Mail Code 661, Greenbelt, MD 20771, USA\\
  $^{10}$Department of Physics, University of California Berkeley, 366 LeConte Hall MC 7300, Berkeley, CA, 94720-7300, USA\\
  $^{11}$Physics Division, Lawrence Berkeley National Laboratory, 1 Cyclotron Road, Berkeley, CA 94720, USA\\
  $^{12}$Carnegie Observatories, Las Campanas Observatory, Casilla 601, La Serena, Chile\\
  $^{13}$Division of Physics, Mathematics, and Astronomy, California Institute of Technology, Pasadena, CA 91125, USA\\
  $^{14}$Spitzer Science Center, California Institute of Technology,  M/S 314-6, Pasadena, CA 91125, U.S.A.\\
  $^{15}$Australian Astronomical Observatory, PO Box 915, North Ryde NSW 1670, Australia\\
  $^{16}$Space Sciences Laboratory, University of California Berkeley, 7 Gauss Way, Berkeley, CA 94720, USA\\
  $^{17}$Department of Physics, Yale University, PO Box 208120, New Haven, CT 06520-8120, USA\\
  $^{18}$Dark Cosmology Centre, Niels Bohr Institute, University of Copenhagen, Juliane Maries Vej 30, 2100 K\o benhavn \O, Denmark\\
  $^{19}$Joint Space Science Institute, University of Maryland, College Park, Maryland 20742, USA\\
}

\begin{document}

\maketitle

\label{firstpage}
\clearpage
\begin{abstract}
We present spectroscopic observations of the host
galaxies of 82 low-redshift type Ia supernovae (SNe Ia) discovered by
the Palomar Transient Factory (PTF). We determine star-formation
rates, gas-phase/stellar metallicities, and stellar masses and ages of
these objects. As expected, strong correlations between the SN Ia
light-curve width (stretch) and the host age/mass/metallicity are
found: fainter, faster-declining events tend to be hosted by
older/massive/metal-rich galaxies. There is some evidence that redder
SNe Ia explode in higher metallicity galaxies, but we found no
relation between the SN colour and host galaxy extinction based on the
Balmer decrement, suggesting that the colour variation of these SNe
does not primarily arise from this source. SNe Ia in
higher-mass/metallicity galaxies also appear brighter after
stretch/colour corrections than their counterparts in lower mass
hosts, and the stronger correlation is with gas-phase metallicity
suggesting this may be the more important variable.  We also compared
the host stellar mass distribution to that in galaxy targeted SN
surveys and the high-redshift untargeted Supernova Legacy Survey
(SNLS). SNLS has many more low mass galaxies, while the targeted
searches have fewer. This can be explained by an evolution in the
galaxy stellar mass function, coupled with a SN delay-time
distribution proportional to $t^{-1}$. Finally, we found no
significant difference in the mass--metallicity relation of our SN Ia
hosts compared to field galaxies, suggesting any metallicity effect on
the SN Ia rate is small.
\end{abstract}


\section{Introduction}
\label{sec:introduction}

Type Ia supernovae (SNe Ia) are remarkable cosmological standardisable
candles that are routinely used to measure cosmological parameters
\citep{1998AJ....116.1009R,1999ApJ...517..565P,2007ApJ...659...98R,2009ApJS..185...32K,2011ApJ...737..102S,2012ApJ...746...85S}.
As these studies become increasingly more precise, systematic
uncertainties become a significant component of the error budget
\citep{2011ApJS..192....1C}. Thus an important consideration in their
future use is the degree to which SN Ia properties evolve with
redshift or depend on their environment -- and how well any
evolutionary effects can be controlled.

The host galaxies and environments of SNe Ia has long been a profitable
route to probe astrophysical effects in the SN Ia population, with the
observed properties of SNe Ia known to correlate with the physical
parameters of their host galaxy stellar populations. SNe Ia in
elliptical or passively evolving galaxies are intrinsically fainter
than SNe Ia in spiral or star-forming galaxies, and possess narrower,
faster evolving (or lower `stretch') light curves
\citep{1995AJ....109....1H,1996AJ....112.2391H,1999AJ....117..707R,2000AJ....120.1479H,2001ApJ...554L.193H,2005ApJ...634..210G,2006ApJ...648..868S}.
The impact of these effects on the cosmological results is small due
to observed correlations between SN Ia light curve shape and
luminosity \citep{1993ApJ...413L.105P}, and between SN Ia optical
colour and luminosity \citep{1996ApJ...473...88R,1998A&A...331..815T}.
When these empirical relations are applied to SN Ia datasets, only
small correlations remain between SN Ia luminosity and host galaxy
properties, such as their stellar masses or star formation rates
\citep{2010ApJ...715..743K,2010MNRAS.406..782S,2010ApJ...722..566L}.

These residual trends between SN luminosity and host galaxy properties
can be accounted for at the level required by current cosmological
analyses, either by directly using host galaxy information in the
cosmological fits \citep{2011ApJ...737..102S} or by applying
probabilistic corrections to the absolute magnitudes of the SNe
\citep{2012ApJ...746...85S}.  However, as the size of other systematic
uncertainties in the cosmological analyses are reduced as, for
example, the accuracy of the photometric calibration procedures
improve \citep[][]{2013A&A...552A.124B}, understanding the physical
origin of these astrophysical correlations will become critical for
future, larger samples \citep[e.g. Dark Energy
Survey;][]{2012ApJ...753..152B}. 

The two primary competing ideas are that either progenitor metallicity
or progenitor age (or a combination of both) play a role in
controlling SN Ia luminosities -- but directly measuring either is
extremely difficult.  Indirect information can be obtained on
metallicity from the ultraviolet (UV) SN spectra
\citep[e.g.][]{1998ApJ...495..617H,2000ApJ...530..966L}, and while
this has provided useful insights into evolution within SN Ia
populations
\citep{2008ApJ...674...51E,2008ApJ...684...68F,2012MNRAS.426.2359M,2012AJ....143..113F},
the interpretation of any individual event is extremely complex even
with very high quality data \citep{2013arXiv1305.2356M}. There is
currently no technique to estimate the age of the progenitor star from
the SN spectrum.

Thus many studies have instead focused on detailed spectroscopic
studies of the host galaxies of the SNe Ia rather than the events
themselves, assembling statistical samples with which to search for
correlations between the physical parameters defining the host
galaxies, and the SN Ia properties. Such global host galaxy properties
are believed to represent reasonable tracers of the SN progenitor
star, at least in a statistical sense \citep{2011MNRAS.414.1592B}.
Common spectroscopic measurements include star formation rates and gas
phase metallicity measured from nebular emission lines
\citep{2005ApJ...634..210G,2011ApJ...743..172D,2012A&A...545A..58S,2012arXiv1211.1386J,2013arXiv1304.4720C,2013arXiv1309.1182R},
and stellar metallicity and age measured from
spectral absorption indices
\citep{2008ApJ...685..752G,2012A&A...545A..58S,2012arXiv1211.1386J}.

A number of intriguing results have arisen from these studies. Based
on star formation activity in the host galaxy, brighter SNe Ia were
found to explode in more active galaxies than those in passive
galaxies.  The SN Ia luminosities were also found to be significantly
correlated with host gas-phase metallicities, with metal-rich galaxies tending
to host fainter SNe Ia than metal-poor galaxies. A similar trend has
also been identified with host stellar metallicity. The stellar age of
the host galaxies also shows a correlation with SN Ia luminosities, in
the sense that fainter SNe Ia preferentially explode in older
populations.

In this paper, we present new spectroscopic observations of the host
galaxies of SNe Ia discovered by the Palomar Transient Factory
\citep[PTF;][]{2009PASP..121.1334R,2009PASP..121.1395L}, a project
designed to explore the optical transient sky. 82 high-quality spectra
of SNe Ia host galaxies were obtained, with precise determinations of
their stellar masses, gas-phase and stellar metallicities, stellar
ages, and star formation rates. We then combine these host parameters
with optical multi-colour light curves of the SNe in an effort to
investigate the physical origin of the trends discussed above.

A plan of the paper follows. In Section~\ref{sec:observ-data-reduct}
we introduce the SN Ia sample, and the spectroscopic observations of
their host galaxies.  Section~\ref{sec:host-galaxy-param} discusses
the various measurements that can be made from these host galaxy
spectra, and the methods for measuring star formation rates, host
galaxy stellar masses, ages and metallicities. In Section
\ref{sec:depend-sn-prop} we examine how the key SN Ia photometric
properties depend on these host parameters, and we discuss our
findings in Section~\ref{sec:discussion}. We conclude in Section
\ref{sec:conclusions}. Throughout this paper, we assume
$\mathrm{H_0}=70$\,km\,s$^{-1}$\,Mpc$^{-1}$ and a flat universe with
$\omatter=0.3$.

\section{OBSERVATIONS AND DATA REDUCTION}
\label{sec:observ-data-reduct}
\begin{figure*}
	\centering
		\includegraphics*[scale=0.8]{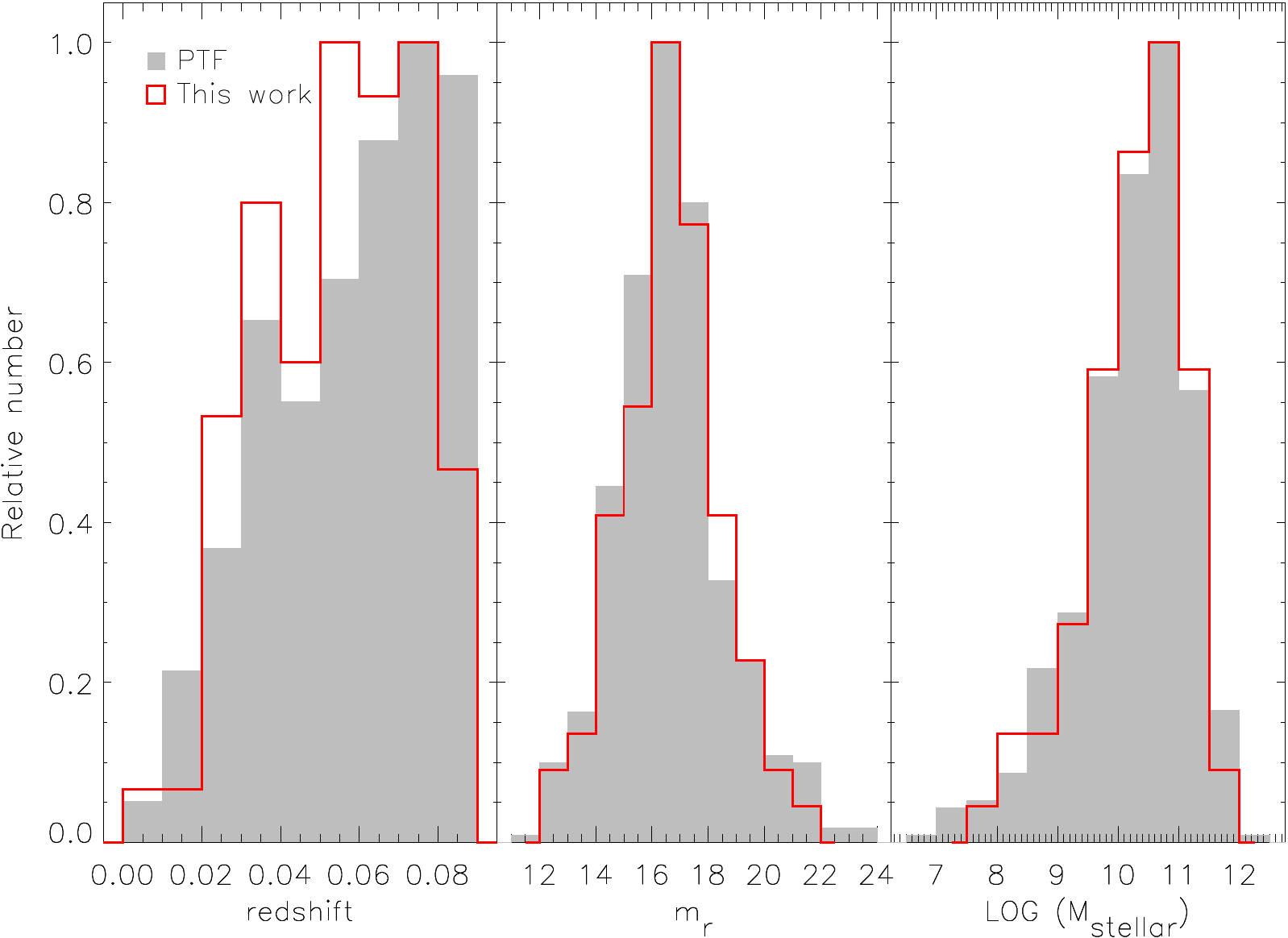}
                \caption{The distribution in SN redshift, SDSS
                  $r$-band host galaxy apparent magnitude
                  ($m_{r}$), and host galaxy stellar mass (\mstellar)
                  of our 82 PTF SNe Ia host galaxies.  The larger PTF
                  SN Ia sample is shown as the filled grey histogram
                  (527 SNe Ia in the redshift histogram, and 443
                  events in the $m_r$ and \mstellar\ panels), and our
                  host galaxy sample studied as the open red
                  histogram.}
        \label{sample_selection}
\end{figure*}

In this section, we present the sample of SNe Ia and their host
galaxies studied in this paper. We discuss the SN sample selection,
the observations of the host galaxies and their data reduction, and
the photometric light curve data for the SNe.

\subsection{SN sample selection}
\label{sec:sample-selection}
The SNe Ia studied in this paper were discovered by the PTF, a project
which operated from 2009-2012 and used the CFH12k wide-field survey camera
\citep{2008SPIE.7014E.163R} mounted on the Samuel Oschin 48-inch telescope (P48)
at the Palomar Observatory. The observational cadences used to discover the
SNe ranged from hours up to $\sim5$ days. SN candidates were identified in
image subtraction data and ranked using both simple cuts on the detection
parameters and a machine learning algorithm \citep{2012PASP..124.1175B}, and then
visually confirmed by members of the PTF collaboration or, from mid-2010 onwards,
via the citizen science project `Galaxy Zoo: Supernova' \citep{2011MNRAS.412.1309S}.
The latter identified 8 of the SNe studied in this paper.

Promising SN candidates were then sent for spectroscopic confirmation using 
a variety of telescope/instrument combinations. These included:
The William Herschel Telescope (WHT) and the Intermediate dispersion Spectrograph and Image System (ISIS),
the Palomar Observatory Hale 200-in and the double spectrograph,
the Keck-I telescope and the Low Resolution Imaging Spectrometer (LRIS),
the Keck-II telescope and the DEep Imaging Multi-Object Spectrograph (DEIMOS),
the Gemini-N telescope and the Gemini Multi-Object Spectrograph (GMOS),
the Very Large Telescope and X-Shooter, the Lick Observatory 3m Shane telescope and the Kast Dual Channel Spectrograph,
the Kitt Peak National Observatory 4m telescope and the Richey-Chretien Spectrograph,
and the University of Hawaii 88-in and the Supernova Integral Field Spectrograph (SNIFS).
All of the spectra used to confirm the SNe in this paper as SN Ia are available from
the WISeREP archive \citep{2012PASP..124..668Y}.

PTF operated in either the $R$ or $g^\prime$ band (hereafter $R_\textrm{P48}$ and $g_\textrm{P48}$),
switching from $g_\textrm{P48}$ band around new moon to $R_\textrm{P48}$ band when the sky was brighter.
Multi-colour light curves were not obtained by default for all SNe using the P48; instead they were
assembled via triggered observations on other robotic facilities, e.g., the Liverpool Telescope \citep[LT;][]{2004SPIE.5489..679S},
the Palomar 60-in \citep[P60;][]{2006PASP..118.1396C} and the Las Cumbres Observatory Global Telescope Network \citep[LCOGT;][]{2013arXiv1305.2437B} Faulkes Telescopes (FTs; clones of the LT).

The full PTF SN Ia sample comprises some 1250 spectroscopically
confirmed events. However, many of these are at relatively high
redshift and thus have poor quality P48 light curves, or were
discovered at the start or end of an observing season and thus have
incomplete P48 light curves. In both of these cases no multi-colour
information is available. Thus the first task is to define a parent
sample of high-quality SNe Ia from which targets for host galaxy
studies can be selected. Several criteria were used. 

Firstly, the PTF SN Ia program (generally) restricted multi-colour
follow-up to those events with a redshift ($z$) of $z<0.09$. The
motivation for this was to define a sample less susceptible to
selection effects: the median redshift of all PTF SNe Ia is 0.1, and
at $z=0.09$, a typical SN Ia has a peak apparent magnitude of
$R_\textrm{P48}\simeq18.5$, $\simeq2.5$\,mag above the PTF detection limit of 21
(a typical SN Ia at $z=0.09$ has $R_\textrm{P48}=21$ at 13 days before maximum light).
We apply the same redshift constraint, giving a parent sample of 527
SNe Ia. Secondly, for this host galaxy study we only considered SNe Ia
with a multi-colour light curve: only SNe Ia discovered and confirmed
before maximum light were sent for detailed monitoring, with around
220 events followed in this way. Finally, for this paper, we only
selected `older' SNe Ia for study, i.e., those SNe Ia which had
already faded by the time the host galaxy spectrum was taken. We took
these at $>$1 year since the SN explosion.  This leaves a potential
sample of 140 events, all discovered during 2009--2011, which are
suitable for our study. Of these events, we had sufficient telescope
time to observe 82 host galaxy spectra, selected at random from the
parent sample.  The host galaxies of the SNe Ia were identified by
inspecting images taken by the SDSS.  Most of the host galaxies in our
sample can be identified unambiguously, except PTF09dav where its
likely host galaxy lies $\sim41$\,kpc from the SN
\citep{2011ApJ...732..118S}.

A final caveat is that any biases that exist in the selection of the parent
PTF sample will also be present in our SN Ia sample. The potentially most
serious of these is the difficulty in finding SNe on very bright galaxy
backgrounds, where the contrast of the SN over the host galaxy is low.
This can occur in the cores of galaxies (e.g., \citet{1979A&A....76..188S})
but also more generally for faint events in bright host galaxies (e.g., \citet{2010AJ....140..518P}),
which of course are also likely to be the most metal rich. However, with
modern image subtraction techniques this is only an issue when the SN
brightness drops to $<10\%$ of that of the host background \citep{2010AJ....140..518P},
and the redshift cuts used in our sample definition mean this is unlikely to occur for normal SNe Ia.

Fig.~\ref{sample_selection} shows a comparison of the distributions of
our host galaxy sample and the various larger PTF samples in redshift,
host galaxy $r$-band apparent magnitude ($m_r$), and host galaxy
stellar mass, \mstellar\ (the determination of \mstellar\ is described
in Section~\ref{sec:host-galaxy-param}). The parent PTF sample shown
in Fig.~\ref{sample_selection} contains the 527 $z<0.09$ PTF SNe Ia,
although only 443 of these have Sloan Digital Sky Survey (SDSS)
$u g r i z$ imaging data from which
\mstellar\ estimates could be made
(Section~\ref{sec:host-photometry}). Of the 84 events for which SDSS
photometry is not available, 74 lie outside the SDSS footprint, and
the remaining 10 SNe Ia have no host galaxy visible in the SDSS
images.  A K-S test gives a 35, 77 and 99 percent probability that
our host galaxy sample and the larger PTF sample are drawn from the
same population in redshift, $m_r$, and \mstellar. Thus we find no
strong evidence that our SN Ia host galaxy sample is biased with
respect to the larger PTF sample.

\subsection{Host galaxy observations}
\label{sec:observations}
\begin{table*}
\centering
\caption{The instrumental setups used for the spectroscopic data.}
\begin{tabular}{cccccc}
\hline\hline
Telescope & Spectrograph & \multicolumn{2}{c}{Gratings/Grisms} & Dichroic & $\lambda$ coverage\\
   &   & (Red) & (Blue) &  & (\AA)\\
\hline
Gemini & GMOS & R400 & B600 & -- & $3600$--$9400$\\
WHT & ISIS & R158R & R300B & 5336\,\AA & $3000$--$10000$\\
Lick & Kast & 300/7500 & 600/4310 & 5500\,\AA & $3000$--$11000$\\
Keck & LRIS & 400/8500 & 600/4000 & 5696\,\AA & $3200$--$10000$\\
\hline
\end{tabular}
\label{setup}
\end{table*}

All of our host galaxy spectra were obtained using spectrographs
operating in long-slit mode on four different facilities.
Table~\ref{setup} summarises the instruments and setups used for our
spectroscopic data, and an observational log of the galaxies studied
in this paper can be found in Table~\ref{obs-log}. Generally, our
strategy was to place the slit through both the positions of the SN
and the centre of the host galaxy. Thus we were careful to ensure that
the observations were taken at low airmass to avoid losses due to not
observing at the parallactic angle. The median airmass of the spectroscopic
data in this work is $\sim1.15$.

Most of our SN Ia host galaxy spectra were taken at the Gemini
Observatory during 2010--2012 (59 out of 82 hosts), using both Gemini
North and Gemini South.  We used GMOS \citep{2004PASP..116..425H} with
a $3600$--$9400$\,\AA\ wavelength coverage provided using two
different settings (B600 and R400 gratings). Two exposures in each
setting were taken, with a $\sim100$ pixel shift in wavelength space
in order to avoid the gaps between the detectors (the GMOS array is
composed of three CCDs). Total integration times were around two hours
per source.

18 further SN Ia host spectra were taken at the 4.2-m WHT using ISIS,
providing $3000$--$10000$\,\AA\ wavelength coverage. ISIS is a
dual-armed spectrograph, and we used the R300B and R158R gratings in
the blue and red arms, respectively. The 5300 dichroic was used.

Two brighter host galaxy spectra were taken with the 3-m Shane
telescope at the Lick Observatory, using the Kast Spectrograph \citep{Kast_spectrograph}
providing $3000$--$11000$\,\AA\ wavelength coverage. Here, the 300/7500
grating was used for the red arm and 600/4310 grism for the blue arm,
using the D55 dichroic.

Finally the 10-m telescope Keck-I telescope was used to observe three
fainter ($m_r\geq20$) host galaxies using LRIS
\citep{1995PASP..107..375O} with a $3200$--$10000$\,\AA\ wavelength
coverage. LRIS is also a dual-armed spectrograph. The 400/8500 grating
was used for the red arm and the 600/4000 grism for the blue arm, with
the D560 dichroic.

\subsection{Spectral data reduction}
\label{sec:data-reduction}
We reduced our data using a custom data reduction pipeline written in
\textsc{iraf}\footnote{The Image Reduction and Analysis Facility
  (\textsc{iraf}) is distributed by the National Optical Astronomy
  Observatories, which are operated by the Association of Universities
  for Research in Astronomy, Inc., under cooperative agreement with
  the National Science Foundation.}.  For data taken at the Gemini
Observatory, we also used some tasks from the Gemini \textsc{iraf}
package. Our pipeline follows standard procedures, including bias
subtraction, flat-fielding, cosmic-ray removal \citep[using
\textsc{lacosmic};][]{2001PASP..113.1420V} and a wavelength
calibration. The \textsc{iraf} task \textsc{apall} is then used to
extract the 1-D spectrum from each 2-D frame, and a (relative) flux
calibration performed with a telluric-correction by comparing to
standard stars.  `Error' spectra are derived from a knowledge of the
CCD properties and Poisson statistics, and are tracked throughout the
reduction procedure. As all the spectra in our sample are taken either
with a spectrograph with two different grating settings (Gemini), or
with dual-arm spectrographs with a dichroic (WHT, Lick, Keck), red and
blue spectra for each object, with different wavelength coverages and
dispersions, need to be combined to produce the final spectrum.  This
was performed by rebinning to a common dispersion, and combining (with
weighting) to form a final contiguous spectrum.

\begin{figure}
	\centering
		\includegraphics*[scale=0.5]{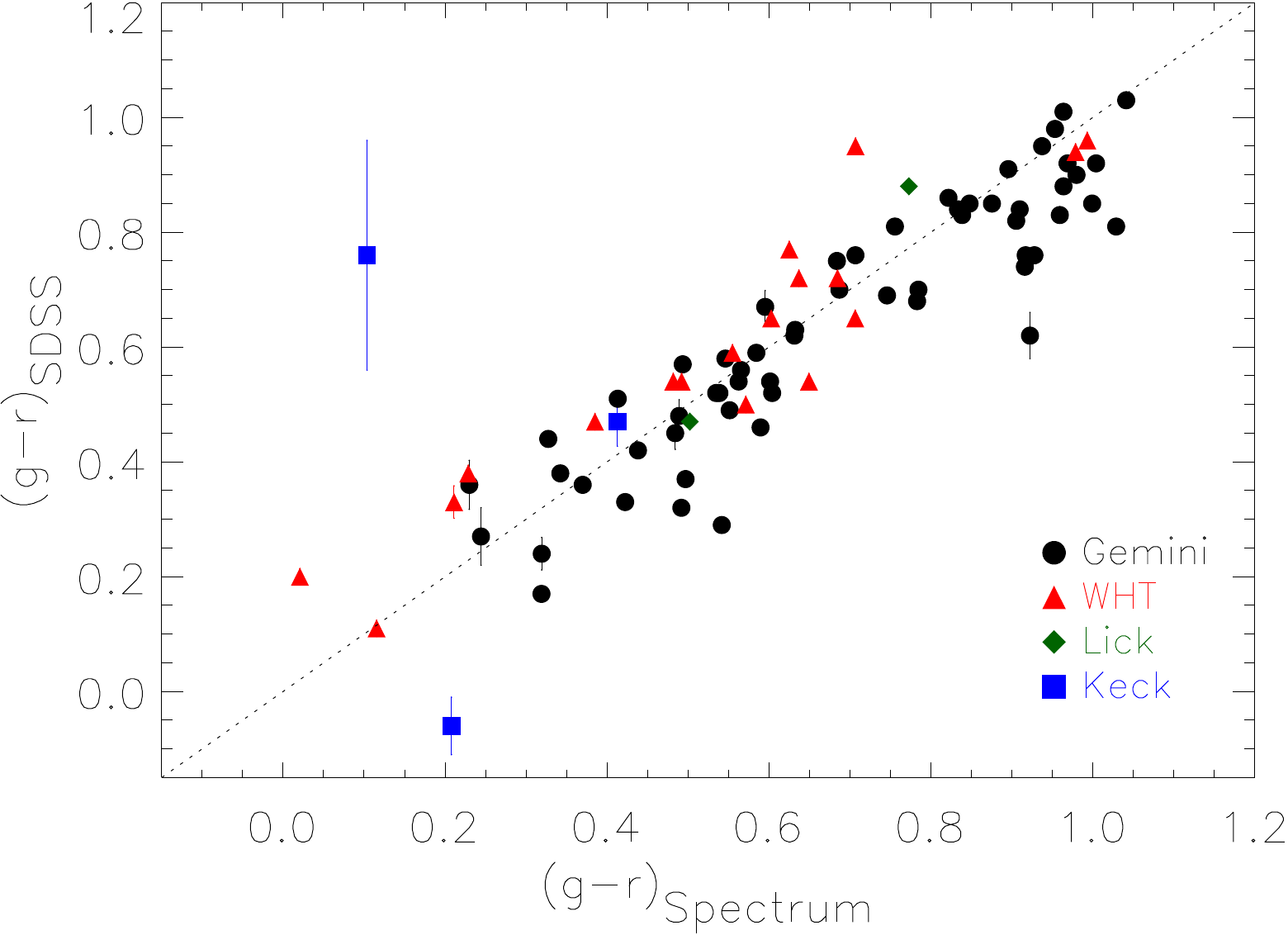}
                \caption{The $g-r$ colour derived from our host galaxy
                  spectra, compared with that determined from the SDSS
                  broad-band photometry. The line of equality is shown
		  in dotted line.}
        \label{g-r}
\end{figure}

\begin{figure*}
	\centering
		\includegraphics*[scale=0.8]{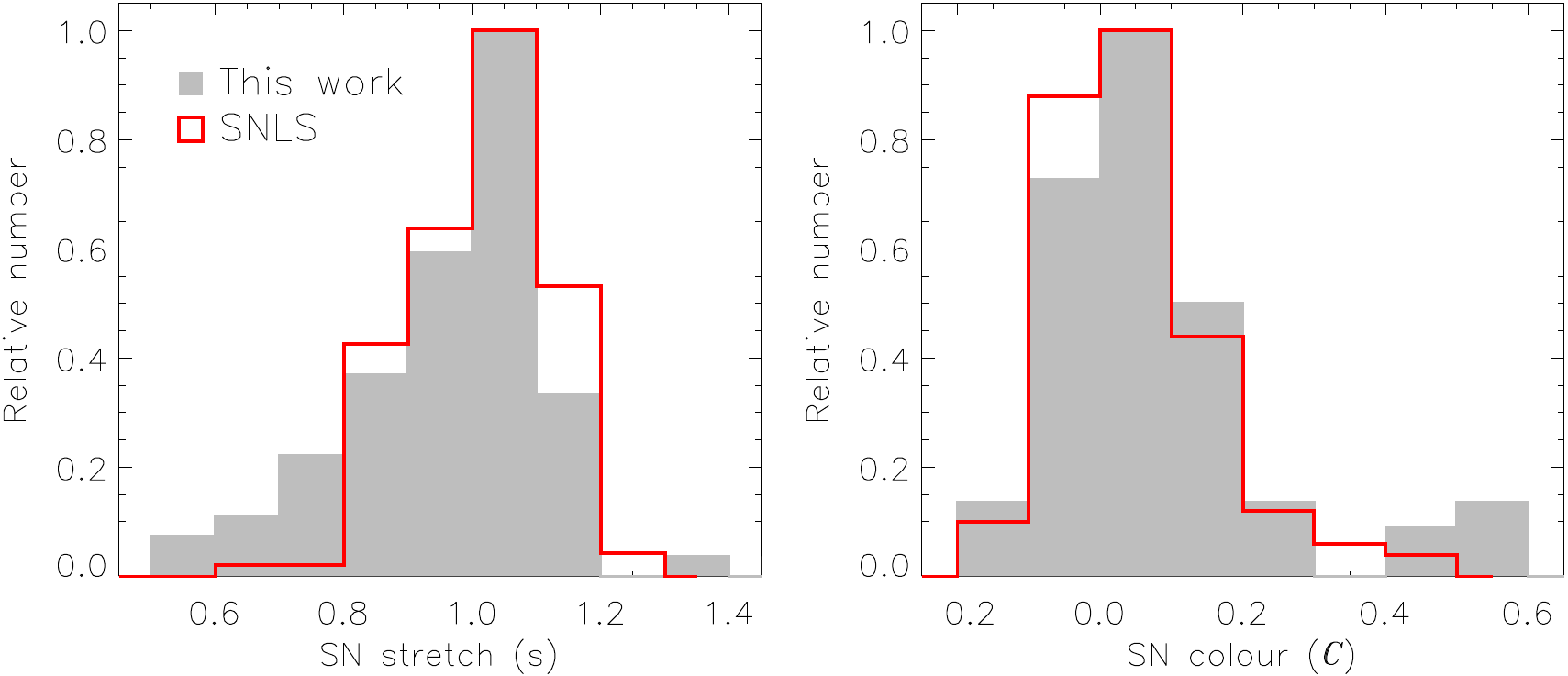}
                \caption{The grey filled histograms show the SN stretch ($s$)
                  and colour (\col) distributions of our PTF sample
                  (see Section~\ref{sec:sn-photometry-light} for more
                  details). The SNLS sample of
                  \citet{2010A&A...523A...7G} at $z<0.6$ is
                  over-plotted in the red open histogram.}
        \label{sn_stretch_colour_statistics}
\end{figure*}

We test our relative flux calibration by comparing synthetic
photometry measured from our final host spectra, with SDSS Data
Release 9 \citep[DR9;][]{2012ApJS..203...21A} photometry of the same
objects. The SDSS model magnitudes are used here.
Fig.~\ref{g-r} shows the $g-r$ colour of our
spectra plotted against the $g-r$ colour from the SDSS
photometry. Overall our data show a good consistency with the SDSS
photometry: the r.m.s. scatter is 0.12\,mag, with a mean offset
of 0.01\,mag.

We correct our absolute flux calibration using the same SDSS
photometry (this is important for host galaxy parameters measured
based on absolute line strength, for example star formation rates).
Again, we measure a synthetic SDSS $r$-band magnitude for our observed
spectra, and compare to the SDSS photometry, scaling our observed
spectra so the two magnitudes are equal.

Finally, we apply a correction for foreground galactic extinction
prior to de-redshifting the spectra into the rest-frame. The latest
calibration \citep{2011ApJ...737..103S} is used, and the typical Milky
Way value $R_V=3.1$ is assumed, using a
\citet*[][CCM]{1989ApJ...345..245C} law.  Although redshift estimates
based on the original SN classification spectrum are available, we
confirm these using emission and absorption lines in the galaxy
spectra; the two redshift measures are consistent in all cases.

The quality of our spectra is quite diverse. We estimate the
signal-to-noise ratio (S/N) over a region in the centre of each
spectrum ($\sim5500-6000$\,\AA). The median flux and standard deviation
within that region are measured, and the S/N taken as the ratio of the
two. Our spectra have a S/N ranging from 5 to 53 with a
median of $\simeq28$.

\subsection{SN photometry and light curve fitting}
\label{sec:sn-photometry-light}
Optical light curves of our SNe Ia in $gri$ were obtained at the LT,
the P60, and the FTs. 
There are 66 events with available LT (54 events), P60 (6 events) or FT
(6 events) light curves, complemented by P48 $R_\textrm{P48}$ (and
sometimes $g_\textrm{P48}$) light curves from the rolling PTF search.
In all cases, reference images were made by stacking data taken $>$1
year after the SN explosion, which was then subtracted from the images
containing SN light to remove the host galaxy. We measure the SN
photometry using a point-spread-function (PSF) fitting method.  In
each image frame, the PSF is determined from nearby field stars, and
this average PSF is then fit at the position of the SN event weighting
each pixel according to Poisson statistics, yielding a SN flux and
flux error.

The SiFTO light curve fitting code \citep{2008ApJ...681..482C} was
used to fit the light curves.  SiFTO works in flux space, manipulating
a model of the spectral energy distribution (SED) and synthesising an
observer-frame light curve from a given spectral time-series in a set
of filters at a given redshift, allowing an arbitrary normalization in
each observed filter (i.e., the absolute colours of the template being
fit are not important and do not influence the fit).  The time-axis of
the template is adjusted by a dimensionless relative `stretch' ($s$)
factor to fit the data, where the input template is defined to have
$s=1$.  Once the observer-frame SiFTO fit is complete, a second step
can be used to estimate rest-frame magnitudes in any given set of
filters, provided there is equivalent observer-frame filter coverage,
and at any epoch.  This is performed by adjusting the template SED at
the required epoch to have the correct observed colours from the SIFTO
fit, correcting for extinction along the line of sight in the Milky
Way, de-redshifting, and integrating the resultant SED through the
required filters. This process is essentially a cross-filter
k-correction, with the advantage that all the observed data contribute
to the SED shape used.

We used SiFTO to determine the time of maximum light in the rest-frame
$B$-band, the stretch, the rest-frame $B$-band apparent magnitude at
maximum light $m_B$, and the $B-V$ colour at $B$-band maximum light,
\col. When estimating the final SN colour via the template SED adjustment,
filters that are very close in effective wavelength can introduce discontinuities
in the adjusted spectrum. Thus we remove the P48 $R_\textrm{P48}$ and $g_\textrm{P48}$
filters in this process where data from the LT, P60, or FTs are also available. Note
that the P48 filters are always used to estimate the stretch and time of maximum light.
Fig.~\ref{sn_stretch_colour_statistics} shows the
distribution of our SNe Ia in stretch and colour. As a comparison, we
over-plot the higher-redshift Supernova Legacy Survey (SNLS) sample
studied by \citet{2010A&A...523A...7G} for SNLS events at $z<0.6$
where the SNLS sample is more complete \citep{2010AJ....140..518P}. We
generally find a good agreement in the stretch and colour
distributions, although our sample probes faster (lower stretch) and
redder SNe Ia than SNLS.

\subsection{Host galaxy photometry}
\label{sec:host-photometry}
In later sections, we will use broad-band photometry of the SN Ia host
galaxies to estimate the host galaxy stellar mass. Where available we
use SDSS $u g r i z$ photometry, but some (five) of our SNe with host
galaxy spectra lie outside the SDSS footprint. For these we instead
use the LT $g^\prime r^\prime i^\prime$ images taken as part of the SN
photometric follow-up campaign, calibrated using observations of
either \citet{2002AJ....123.2121S} standard stars, or of the SDSS
stripe 82 \citep{2007AJ....134..973I}. The host photometry is measured
by \textsc{sextractor} \citep{1996A&AS..117..393B}, which we use in
dual-image mode with FLUX\_AUTO photometry, ensuring the same
consistent aperture is used in each filter.

\section{HOST GALAXY PARAMETER DETERMINATION}
\label{sec:host-galaxy-param}

Having described the sample and data that make up our host galaxy
sample, we now discuss the techniques used to fit the SN Ia host
galaxy spectra, and estimate various physical parameters such as the
star formation rate (SFR) and the gas-phase metallicity. We use
various techniques, including emission line measurements to determine
SFRs and gas-phase metallicities, spectral fitting to determine
stellar metallicities and ages, and broad-band photometric fitting to
determine stellar masses. We first introduce the technique used to
make the emission line measurements.

\subsection{Emission line measurement}
\label{sec:emiss-line-meas}
The emission lines and stellar continuum of the host galaxy spectra
are fit using the Interactive Data Language (\textsc{idl}) codes
\textsc{ppxf} \citep{2004PASP..116..138C} and \textsc{gandalf}
\citep{2006MNRAS.366.1151S}. \textsc{ppxf} fits the line-of-sight
velocity distribution (LOSVD) of the stars in the galaxy in pixel
space using a series of stellar templates. The advantage of working in
pixel space is that emission lines and bad pixels are easily excluded
when fitting the continuum. Before fitting the stellar continuum, a
list of emission lines is used to mask this potential contamination.
The stellar templates are based on the MILES empirical stellar library
\citep{2006MNRAS.371..703S,2010MNRAS.404.1639V}, giving a wavelength
coverage of 3540\,\AA\ to 7410\,\AA\ with a spectral resolution of
2.51\,\AA, and a variety of different metallicities and ages.  A total
of 276 templates are selected with $[M/H]=-1.71$ to $+0.22$ in 6 steps
and ages ranging from $0.079$ to $14.12$\,Gyr in 46 steps.

After measuring the stellar kinematics with \textsc{ppxf}, the
emission lines and stellar continuum are simultaneously fit by
\textsc{gandalf}. \textsc{gandalf} treats the emission lines as additional Gaussian
templates. Through an iterative fitting process, \textsc{gandalf}
locates the best velocities and velocity dispersions of each Gaussian
template and also the optimal combination of the stellar templates
which have already been convolved with the LOSVD. This results in the
emission lines and stellar continuum being fit simultaneously to each
spectrum.

Extinction is handled using a two-component reddening model. The first
component assumes a diffusive dust throughout the whole galaxy that
affects the entire spectrum including emission lines and the stellar
continuum, while the second is a local dust component around the
nebular regions, and therefore affects only the emission lines.  The
first component is determined by comparing the observed spectra to the
un-reddened spectral templates. However, the local dust component is
constrained only if the Balmer decrement (the H$\alpha$ $\lambda6563$
to H$\beta$ $\lambda4861$ line ratio) can be measured.  For galaxies
without Balmer lines in their spectra, only the diffusive dust
component is fit (26 out of 82 hosts). To ensure the emission lines in our spectrum are
well-measured, we required a S/N $>3$ (S/N is defined as the
ratio of line amplitude to the noise of the spectrum) for emission
lines used in the determination of host parameters.

\subsection{AGN Contamination}
\label{sec:agn-contamination}
\begin{figure}
	\centering
		\includegraphics*[scale=0.5]{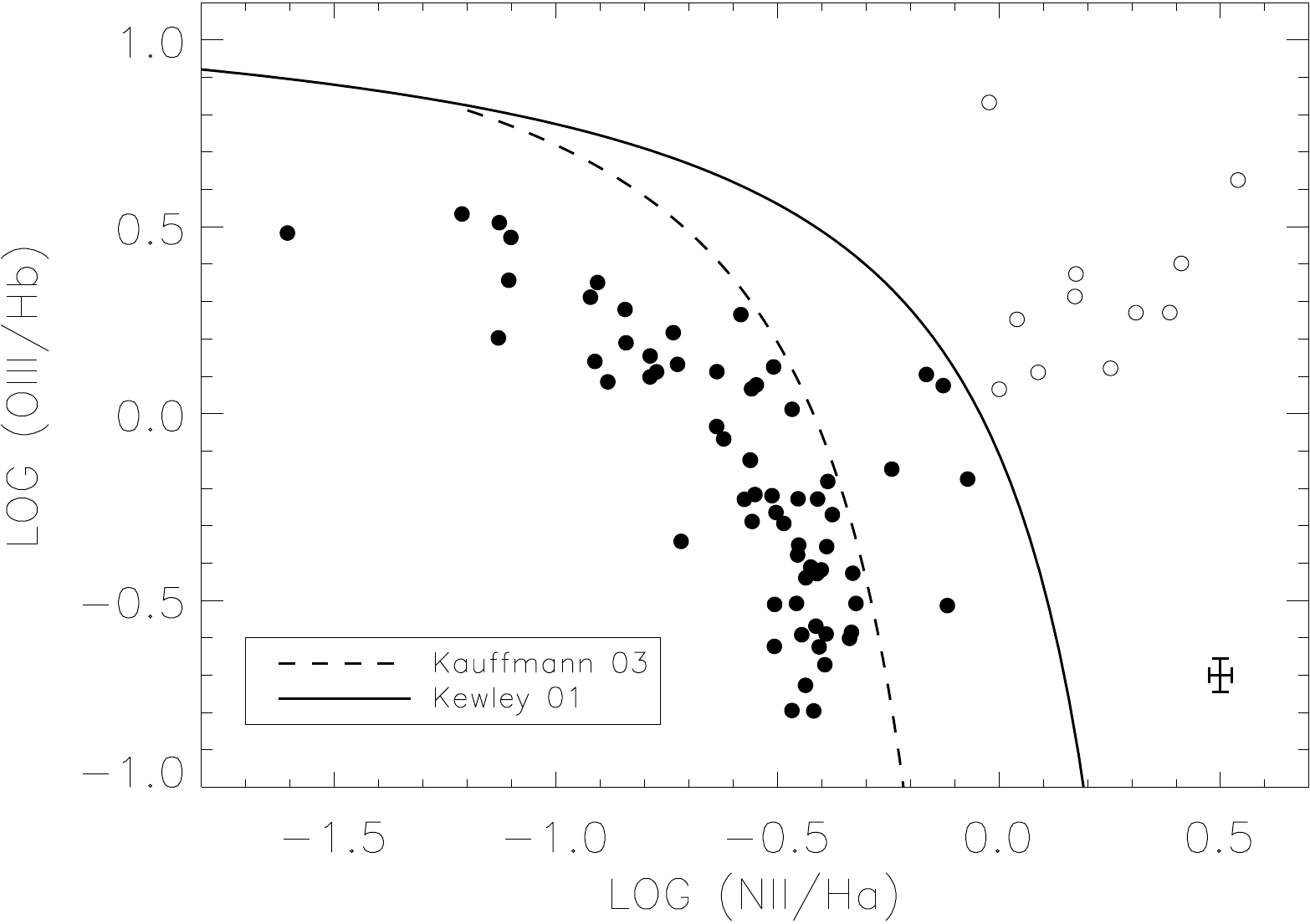}
                \caption{The BPT diagram used to identify the AGN host
                  galaxies in our sample. Two different criteria are
                  over-plotted: \citet{2001ApJ...556..121K} and
                  \citet{2003MNRAS.346.1055K}. The galaxies which lie
                  on the right hand side of Kewley 01 criteria will be
                  regarded as potential AGN host galaxies in this work
                  (the open circles). Normal star-forming galaxies are
                  plotted in filled circles. The representative error is
                  shown in the bottom-right corner.}
        \label{bpt}
\end{figure}

Our next task is to check for active galactic nuclei (AGN) activity in
our host galaxies. In galaxies hosting an AGN, non-thermal emission
from the AGN can dominate over that from the hot stars, leading to a
different ionisation source for the nebular \hii\ regions. This in
turn means that the emission line measurements performed in the
previous section cannot be interpreted using the techniques discussed
later in this section.

We adopt the BPT diagram \citep*{1981PASP...93....5B}, shown in
Fig.~\ref{bpt} for our sample. The galaxies are divided into two
groups using either the criteria proposed by
\citet{2001ApJ...556..121K} or \citet{2003MNRAS.346.1055K}. Any
galaxies lying to the right of these lines in Fig.~\ref{bpt} are
regarded as potential AGN host galaxies. We adopt the
\citeauthor{2001ApJ...556..121K} criterion: a galaxy will be
identified as a AGN if
\begin{equation}
\log\left([\oiii]/\mathrm{H}{\beta}\right)>\frac{0.61}{\log\left([\nii]/\mathrm{H}{\alpha}\right)-0.47}+1.19
\end{equation}
where \nii\ is the flux of $\lambda 6584$ line, and \oiii\ the
$\lambda4959,5007$ lines.  However, this requires the four emission
lines to be well detected.  For those spectra with only \oiii\ and
H$\beta$ or \nii\ and H$\alpha$ available, `two-line' methods can be
used \citep{2003ApJ...597..142M}: a galaxy will be
identified as a AGN if
\begin{equation}
\log\left([\nii]/\mathrm{H}\alpha\right)>-0.2 
\end{equation}
or
\begin{equation}
\log\left([\oiii]/\mathrm{H}\beta\right)>0.8
\end{equation}
Note that these two-line criteria are more conservative than \citeauthor{2001ApJ...556..121K} criterion.
There are 11 (3 by the two-line methods) galaxies in our sample identified as AGN, and these are
discarded from the sample for further emission line analyses. A further 5 galaxies would have been excluded
based on the \citeauthor{2003MNRAS.346.1055K} criterion. We have checked that including these objects does
not affect our results.

\subsection{Determination of host parameters}
\label{sec:determ-host-param}
Having measured the emission lines of the SN hosts, and removed
galaxies likely hosting AGN from our sample, we now turn to the
estimation of various host galaxy physical properties: the host galaxy
SFR, the gas-phase metallicity, the mean stellar metallicity and age,
and the stellar mass. A complete list of the host
parameters measured in this section can be found in
Table~\ref{host_para_phot} and Table~\ref{host_para_spec}.

\begin{figure*}
		\includegraphics*[scale=0.9]{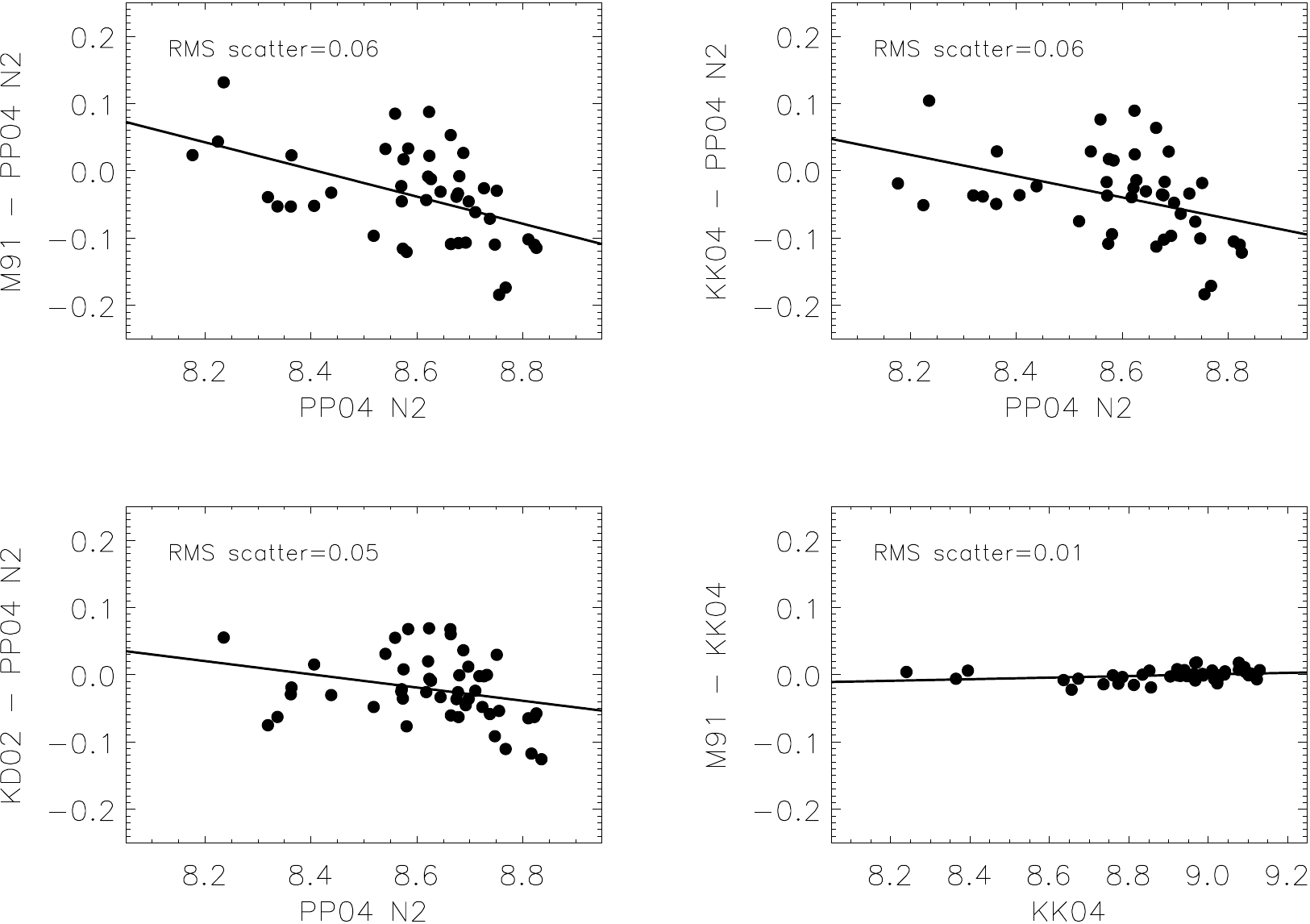}
                \caption{The metallicity conversions used in this
                  paper. We convert the metallicities derived by M91,
                  KK04 and KD02 to
                  \citetalias{2004MNRAS.348L..59P} N2, which is the
                  calibration used for our host galaxy data. In each
                  plot we use the best linear fit (solid line) and
                  scatter to represent the accuracy of the conversion.
                }
        \label{z-z}
\end{figure*}

\subsubsection{Star formation rate}
\label{sec:star-formation-rate}

The SFR of a galaxy can be estimated using nebular lines in the
spectrum, with H$\alpha$ the most popular choice due to its intrinsic
strength and location in the redder part of the spectrum, leading to a
lower susceptibility to dust extinction.  As this emission line is
produced from ionising photons generated by the most massive, youngest
stars, the SFR estimated is a nearly instantaneous measure.  We adopt
the conversion of \citet{1998ARA&A..36..189K}, which used evolutionary
synthesis models to relate the luminosity of the H$\alpha$ line,
$L(\mathrm{H}\alpha)$, to the SFR via
\begin{equation}
  \label{eq:sfr}
  \mathrm{SFR} = 7.9 \times 10^{-42} \times L(\mathrm{H}\alpha)\;M_{\odot}\,\mathrm{yr}^{-1}
\end{equation} 
with $L(\mathrm{H}\alpha)$ measured in erg\,s$^{-1}$. The relation
assumes case B recombination and a \citet{1955ApJ...121..161S}
initial mass function (IMF).  \citet{2004MNRAS.351.1151B} studied the
likelihood distribution of the conversion factor between
$L(\mathrm{H}\alpha)$ and SFR, and found a $\sim0.4$ dex variation
between high-mass and low-mass galaxies, with the
\citet{1998ARA&A..36..189K} conversion factor close to the median
value of their study. As a result, and following
\citet{2011ApJ...743..172D}, we adopt a 0.2\,dex uncertainty in our
SFR measurements.

\subsubsection{Gas-phase metallicity}
\label{sec:gas-phase-metall}

There are various methods for calibrating the the gas-phase
metallicity determined by emission line ratios \citep[for a review
see][hereafter KE08]{2008ApJ...681.1183K}. The direct method is to
measure the ratio of the [\oiii] $\lambda4363$ line to a lower
excitation line to estimate the electron temperature ($T_e$) of the
gas, and then convert it to the metallicity -- the so-called
$T_e$-based metallicity. The disadvantage is that this [\oiii] line is
very weak and difficult to detect unless a very high S/N spectrum can
be acquired; in our sample for only three
spectra was the [\oiii] $\lambda4363$ line detected.
Instead, we use indirect metallicity calibrations, and, following the
recommendation of \citetalias{2008ApJ...681.1183K}, adopt the
empirical relations from \citet[][hereafter
PP04]{2004MNRAS.348L..59P}.  \citetalias{2004MNRAS.348L..59P} fit the
relations between various emission line ratios and the $T_e$-based
metallicity measurement for a sample of \hii\ regions.

The \citetalias{2004MNRAS.348L..59P} `N2' method uses the ratio of
[\nii] $\lambda 6584$ to H$\alpha$.  As these lines are very close in
wavelength space, this is a (nearly) reddening-free method, and covers
both the upper-branch ($\log$([\nii]/[\oii]) $>$$-1.2$) and
lower-branch ($\log$([\nii]/[\oii]) $<$$-1.2$)) metallicities.
The \citetalias{2004MNRAS.348L..59P} relation is only valid for
$-2.5<\mathrm{N}2<-0.3$.  Gas-phase metallicities for 53 galaxies in
our sample can be derived using this method.

For those galaxies outside the valid range of \citetalias{2004MNRAS.348L..59P} `N2' method,
we follow the \citet[][hereafter KD02]{2002ApJS..142...35K}
method. Unlike the empirical \citetalias{2004MNRAS.348L..59P}
calibration, the \citetalias{2002ApJS..142...35K} technique is derived
based on stellar evolution and photoionization models. For the upper
branch metallicities, we use the ratio of [\nii] and [\oii]
$\lambda3727$,
and for the lower-branch metallicity, \citetalias{2002ApJS..142...35K}
recommend averaging two methods based on the $R_{23}$ ratio
($(\oii+\oiii\lambda4959,5007)$/H$\beta$): the \citet[][hereafter
M91]{1991ApJ...380..140M} and \citet[][hereafter
KK04]{2004ApJ...617..240K} relations. For the 29 galaxies where
metallicities are unavailable via the \citetalias{2004MNRAS.348L..59P}
N2 method, 19 can be calibrated following
\citetalias{2002ApJS..142...35K}.  The 10 other galaxies show no
detectable emission lines available for the metallicity calibration.

As previous studies have noted, offsets may exist between these
different metallicity calibrations and thus we used the
self-consistent calibrations of \citetalias{2008ApJ...681.1183K}. For
galaxies where it is possible to make more than one metallicity
measurement, we can also compare the results directly. This is shown in
Fig.~\ref{z-z}; the different calibrations after applying a linear fit compare well.
The observed r.m.s. scatters are (this work/\citetalias{2008ApJ...681.1183K}):
0.06/0.07, 0.06/0.07, 0.05/0.05 and 0.01/0.02 for the
M91--PP04, KK04--PP04, KD02--PP04 and M91--KK04 relations,
respectively. The best-fitting linear trends (the difference between two
different metallicity calibrations) are applied to our metallicity
measurements, although they have no significant effects on the final
results.

\subsubsection{Stellar metallicity and age}

The stellar metallicity and age are normally determined using
absorption features in the spectrum. One widely used method is the
Lick/IDS system \citep{1994ApJS...95..107W,1998ApJS..116....1T}.
Recently, with the availability of high-quality model templates, the
`full spectrum fitting' method has become a popular alternative
\citep[e.g.][]{2005MNRAS.358..363C,2009A&A...501.1269K} to study the
stellar populations, as it exploits more spectral information than
just individual line indices.

In this study we use \textsc{ppxf} to fit the stellar continuum of our
host spectra. The same \textsc{miles} templates described in
Section~\ref{sec:emiss-line-meas} were used. One feature of
\textsc{ppxf} is the linear regularisation performed during the fit,
which can help smooth the weights of the best-fit templates. However,
this feature must be used with caution, as it is a trade-off between
the smoothness and goodness of the fit.  Following the procedure
described in \citet{1992nrfa.book.....P}, the regularisation parameter
for each host galaxy was determined such that the resulting fit was
consistent with the observations, but also gave a smooth star
formation history. Finally, the stellar metallicity and age can be
estimated by performing a weighted-average of all the model
templates, given by
\begin{equation}
\langle\log t\rangle=\sum^{N}_{i=1} w_{i}\times \log t_{i}
\end{equation}
and
\begin{equation}
\langle \mathrm{[M/H]}\rangle=\sum^{N}_{i=1} w_{i}\times \mathrm{[M/H]}_{i},
\end{equation}
where $\log t_{i}$, [M/H]$_{i}$ and $w_{i}$ represent the stellar age,
stellar metallicity and weight of the $i$th template.  The
$\langle\log t\rangle$ and $\langle \mathrm{[M/H]}\rangle$ are the
mass-weighted age and metallicity over the $N$ templates used to fit the
spectrum.  Here we estimated the uncertainty by examining the
dispersion between the results with and without regularization.  An
uncertainty of 0.12\,dex and 0.15\,dex was determined and added to
[M/H] and stellar age, respectively.

\begin{figure}
	\centering
		\includegraphics*[scale=0.5]{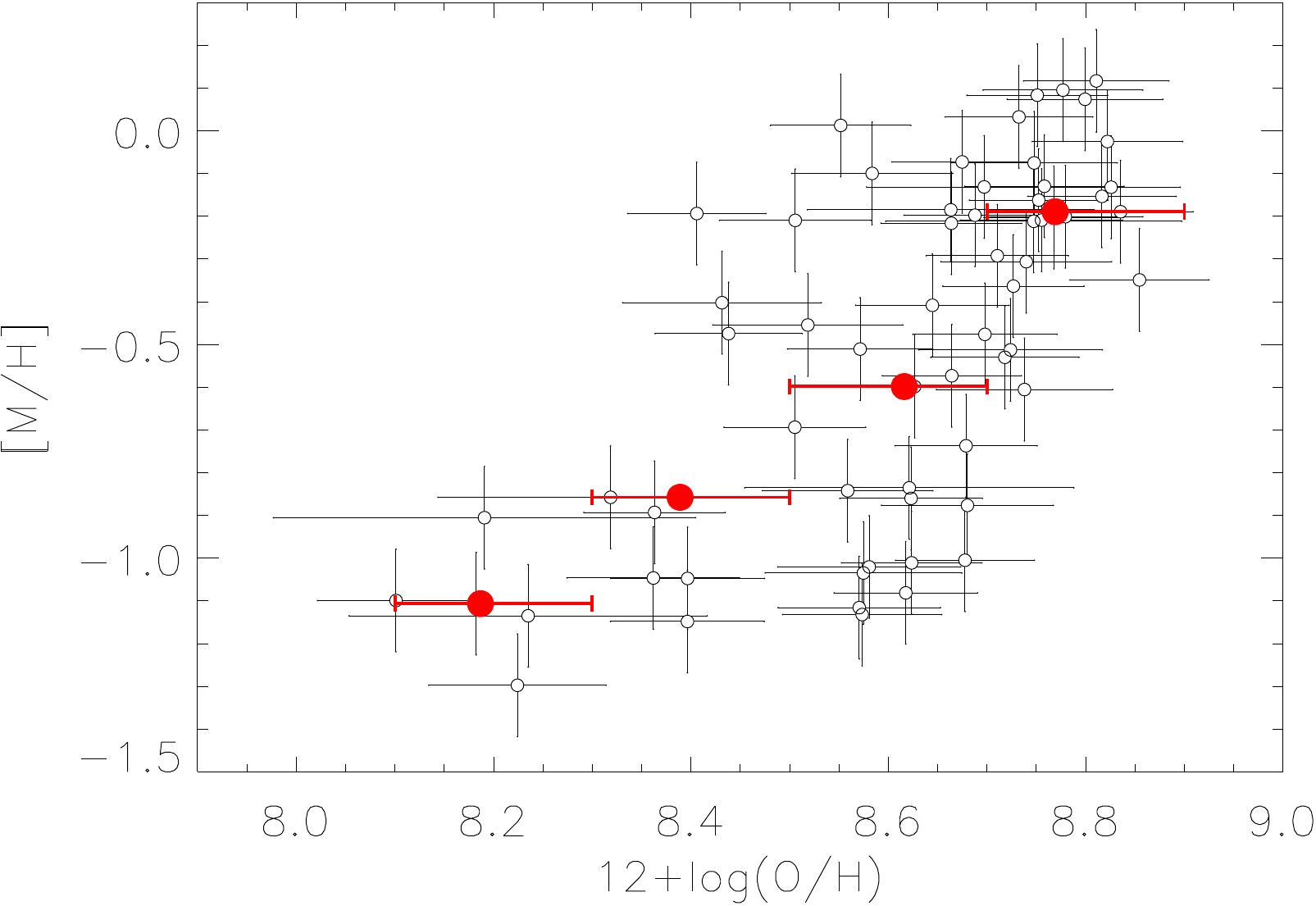}
                \caption{The host stellar metallicity as function of gas-phase
                metallicity. The median metallicity in each bin
                with bin size = 0.2\,dex is computed and shown in red filled-circle.
                }
        \label{sz-gz}
\end{figure}

A comparison between the host gas-phase and stellar metallicities can
be found in Fig.~\ref{sz-gz}. It is clear that the two metallicities
scale with each other with a positive Pearson correlation coefficient
$\sim0.67$.

\subsubsection{Host stellar mass}
\label{sec:host-stellar-mass-1}

The final parameter of interest is the stellar mass of the host
galaxies. We use the photometric redshift code \textsc{z-peg}
\citep{2002A&A...386..446L}, which is based on the spectral synthesis
code P\'{E}GASE.2 \citep{1997A&A...326..950F}, to estimate \mstellar.
\textsc{z-peg} fits the observed galaxy colours
(Section~\ref{sec:host-photometry}) with galaxy SED templates
corresponding to 9 spectral types (SB, Im, Sd, Sc, Sbc, Sb, Sa, S0 and
E). We assume a \citet{1955ApJ...121..161S} IMF.  A foreground dust
screen varying from a colour excess of $E(B-V)=0$ to 0.2\,mag in steps
of 0.02\,mag is used.

\textsc{z-peg} is used to locate the best-fitting SED model (in a
$\chi^2$ sense), with the redshift fixed at the redshift of the SN
host galaxy measured from our spectra.  The current \mstellar\ and the
recent SFR, averaged over the last 0.5\,Gyr before the best fitting
time step, are recorded. Error bars on these parameters are taken from
their range in the set of solutions that have a similar $\chi^2$
\citep[as in][]{2006ApJ...648..868S}. Note that these SFRs are not used
in the analysis in this paper.

\begin{figure*}
		\includegraphics*[scale=0.9]{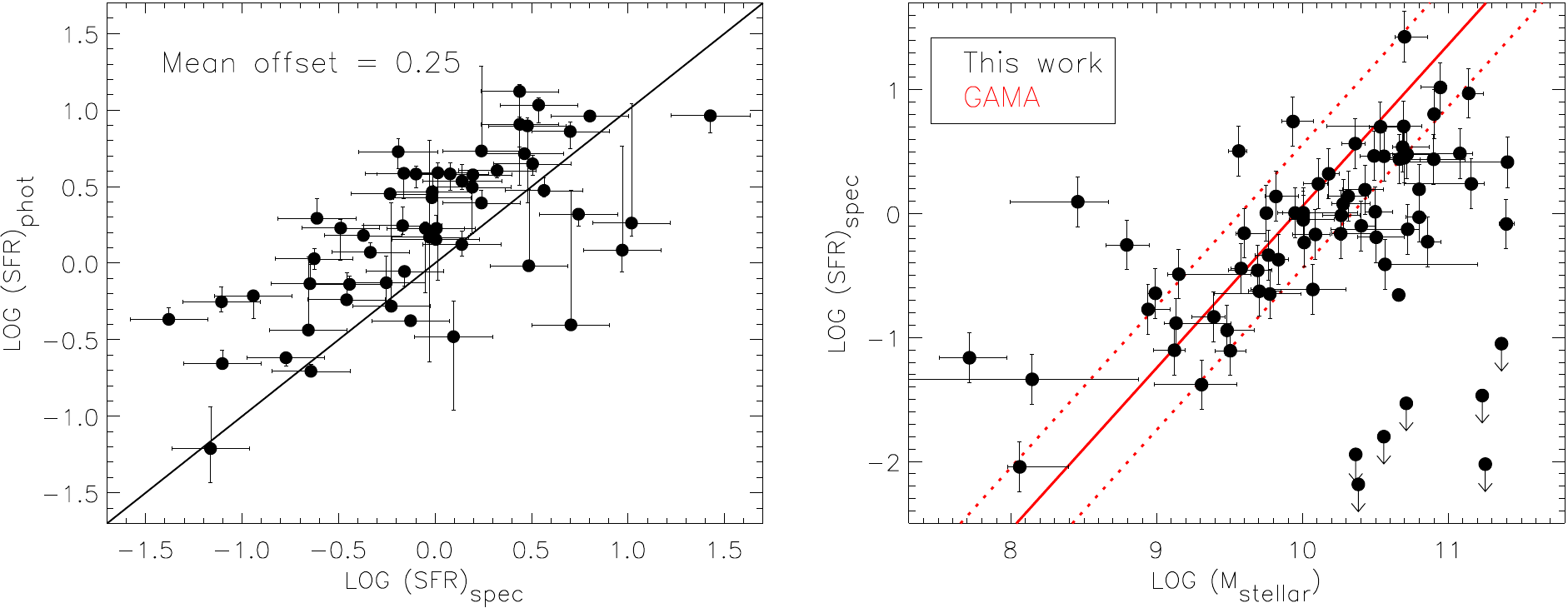}
                \caption{Left: A comparison of the star formation
                  rates (SFRs) derived from H$\alpha$ luminosities
                  (SFR$\mathrm{_{spec}}$) to those measured from
                  broad-band photometry using \textsc{z-peg}
                  (SFR$\mathrm{_{phot}}$). The solid line shows the
                  1:1 relation.  Right: The SFR measured from
                  H$\alpha$ as function of the host \mstellar.  The
                  relation determined by the Galaxy And Mass Assembly
                  survey \citep[GAMA;][]{2012A&A...547A..79F} is
                  over-plotted.  The dotted line shows the 1-$\sigma$
                  range of the GAMA relation. Some passive galaxies
                  with a low SFR for their \mstellar\ (i.e. a low
                  specific SFR) can be seen for large \mstellar. }
        \label{m-sfr}
\end{figure*}

\begin{figure*}
		\includegraphics*[scale=0.8]{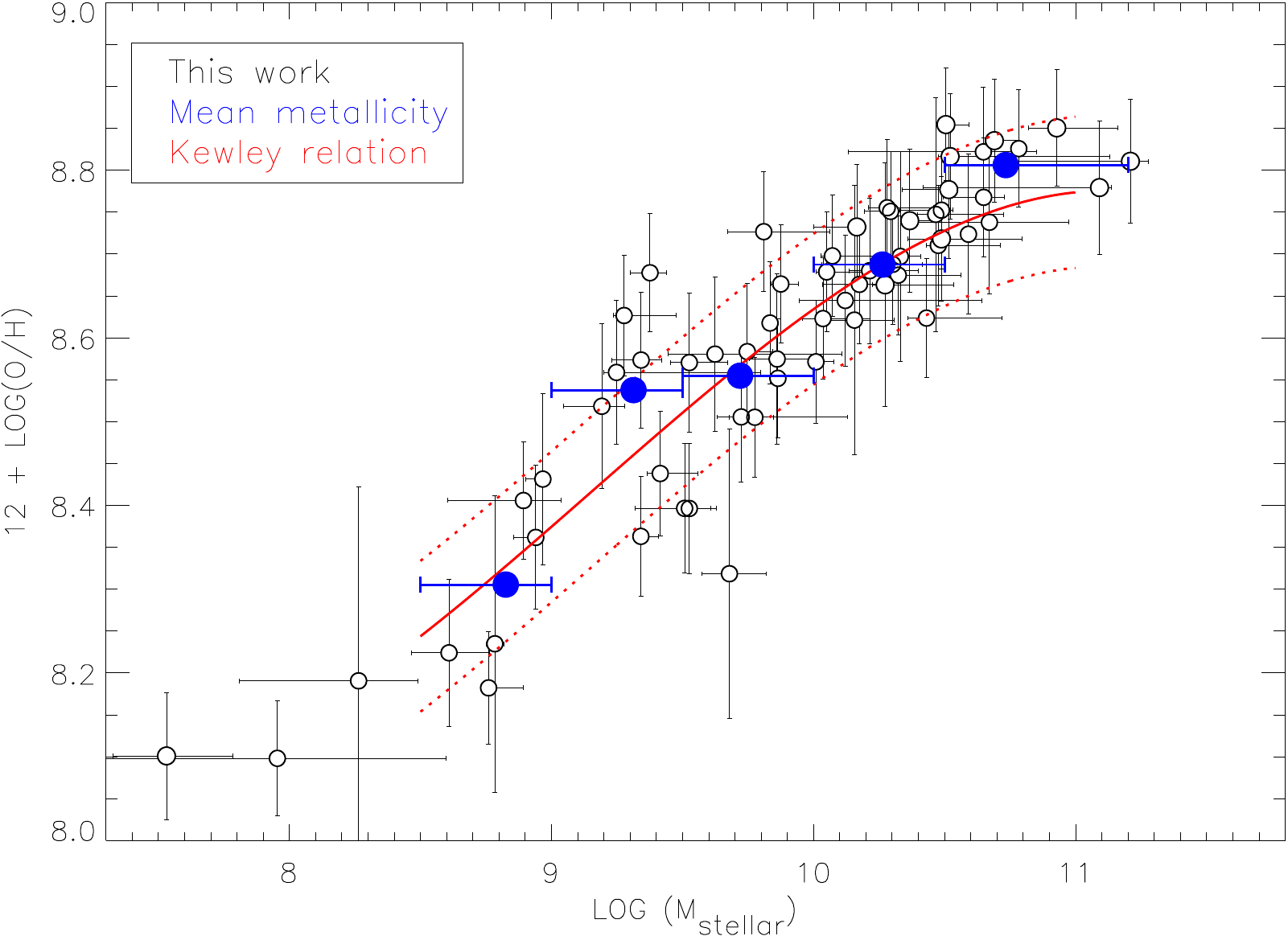}
                \caption{The \mstellar-metallicity relation derived
                  for our host galaxy sample. The red solid line is
                  the best fit from \citetalias{2008ApJ...681.1183K}
                  using the \citetalias{2004MNRAS.348L..59P} N2
                  metallicity calibration. The red-dotted lines
                  represent the r.m.s.  residuals from the best fit
		  to field galaxies. The mean metallicity (blue filled-circle) 
                  in each bin is also computed.}
        \label{m-gz_k01}
\end{figure*}

The main uncertainty in this procedure is the choice of SED libraries
used in the $\chi^2$ fitting. We use the standard \textsc{z-peg}
libraries for ease of comparison to previous results in the
literature. However we note that improved stellar masses can be
obtained by the use of more recent templates
\citep{2012arXiv1211.1386J}, particularly those that include an
improved treatment of the thermally-pulsing Asymptotic Giant Branch
stage of stellar evolution \citep{2005MNRAS.362..799M}.
A fuller discussion of the uncertainties associated with this
stellar population modelling can be found in \cite{2013arXiv1304.4720C}.
These authors conservatively concluded that the maximal systematic is
$\sim 0.4$\,dex in \mstellar, which should be borne in mind when
interpreting our results.

Fig.~\ref{m-sfr} shows a comparison between the SFRs derived from the
H$\alpha$ line to that estimated by \textsc{z-peg}. The mean
difference in $\mathrm{log(SFR)}$ is $\sim0.25$\,dex, with the SFRs
from \textsc{z-peg} systematically larger than those from the
H$\alpha$ luminosity.  A similar offset using similar techniques was
also found by \citet{2012ApJ...755...61S}. This offset is perhaps not
surprising; \textsc{z-peg} determines SFRs essentially from $u$-band
data and is therefore sensitive to SFRs over a longer time-period than
the instantaneous H$\alpha$-based measures.

The relation between the spectroscopic SFRs and \mstellar\ for our
sample is shown in Fig.~\ref{m-sfr}. We over-plot the relation
determined by the Galaxy And Mass Assembly survey
\citep[GAMA;][]{2012A&A...547A..79F}, which used a similar method as
ours for estimating the SFR.  Our results show good consistency with
this relation, although we also sample some massive galaxies with
lower SFRs than the linear relation would predict.

Finally, in Fig.~\ref{m-gz_k01} we plot our metallicities as a
function of \mstellar\ (the `mass--metallicity relation'). The
mass--metallicity relation studied by \citetalias{2008ApJ...681.1183K}
using the \citetalias{2004MNRAS.348L..59P} N2 method is over-plotted
for comparison.  For consistency, in this plot we have adopted the
same IMF \citep{2001MNRAS.322..231K} and cosmology
($\mathrm{H_{0}=72\,km\,s^{-1}\,Mpc^{-1}}$ and $\omatter=0.29$) as
used by \citetalias{2008ApJ...681.1183K} for the measurement of
\mstellar. It is clear that our SN Ia host galaxies follow a very
similar mass--metallicity relation as that of
\citetalias{2008ApJ...681.1183K}. This will be considered further in
Section~\ref{sec:m-z-relation}.

\begin{figure*}
		\includegraphics*[scale=1]{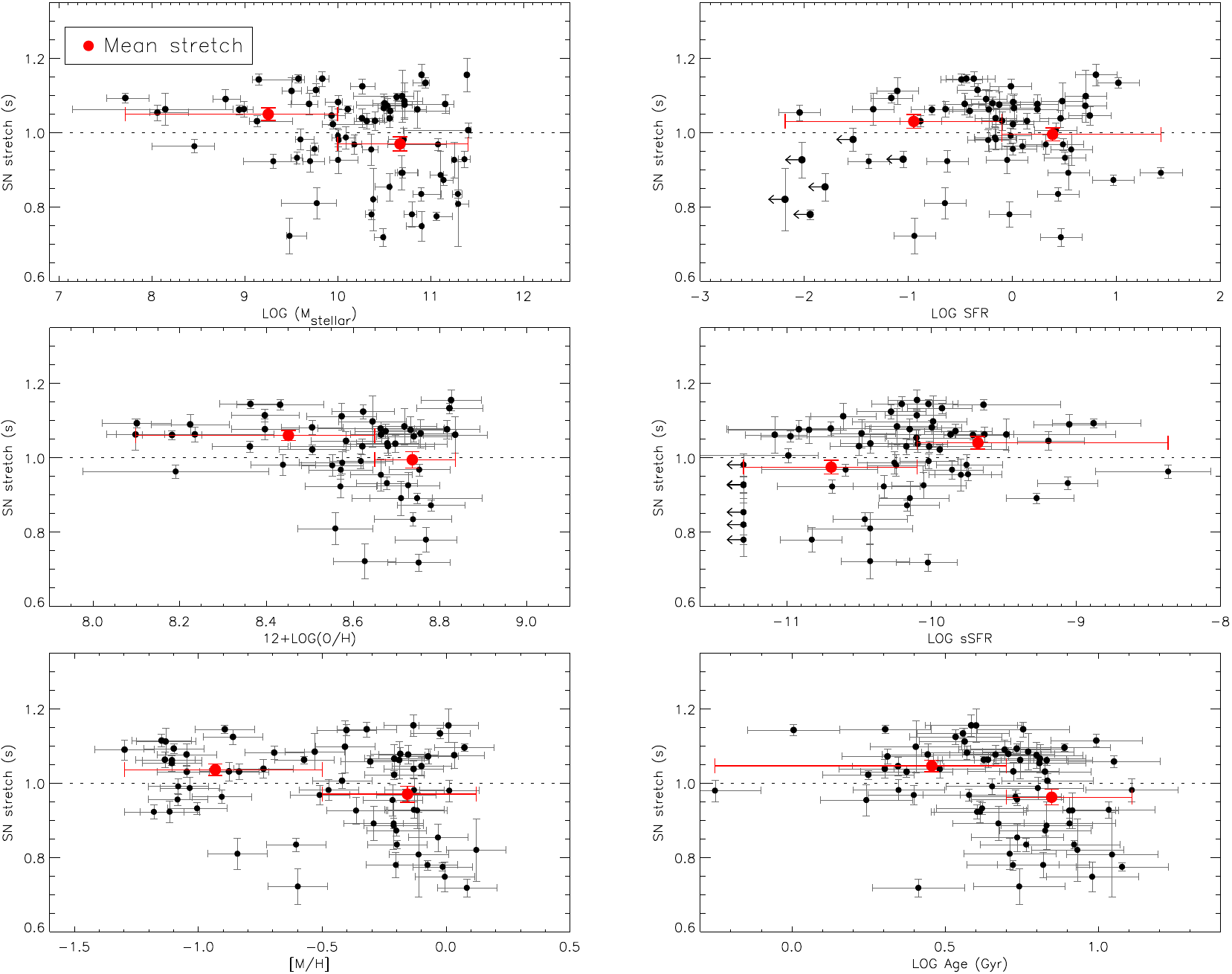}
                \caption{The SN stretch $s$ as a function of host
                  \mstellar\ (top left), SFR (top right), gas-phase
                  metallicity (middle left), specific SFR (sSFR;
                  middle right), mass-weighted mean stellar metallicity (lower left) and
                  stellar age (lower right). The
                  red points represent the mean stretch in bins of
                  host parameters.}
        \label{sn_para1}
\end{figure*}

\begin{figure*}
		\includegraphics*[scale=1]{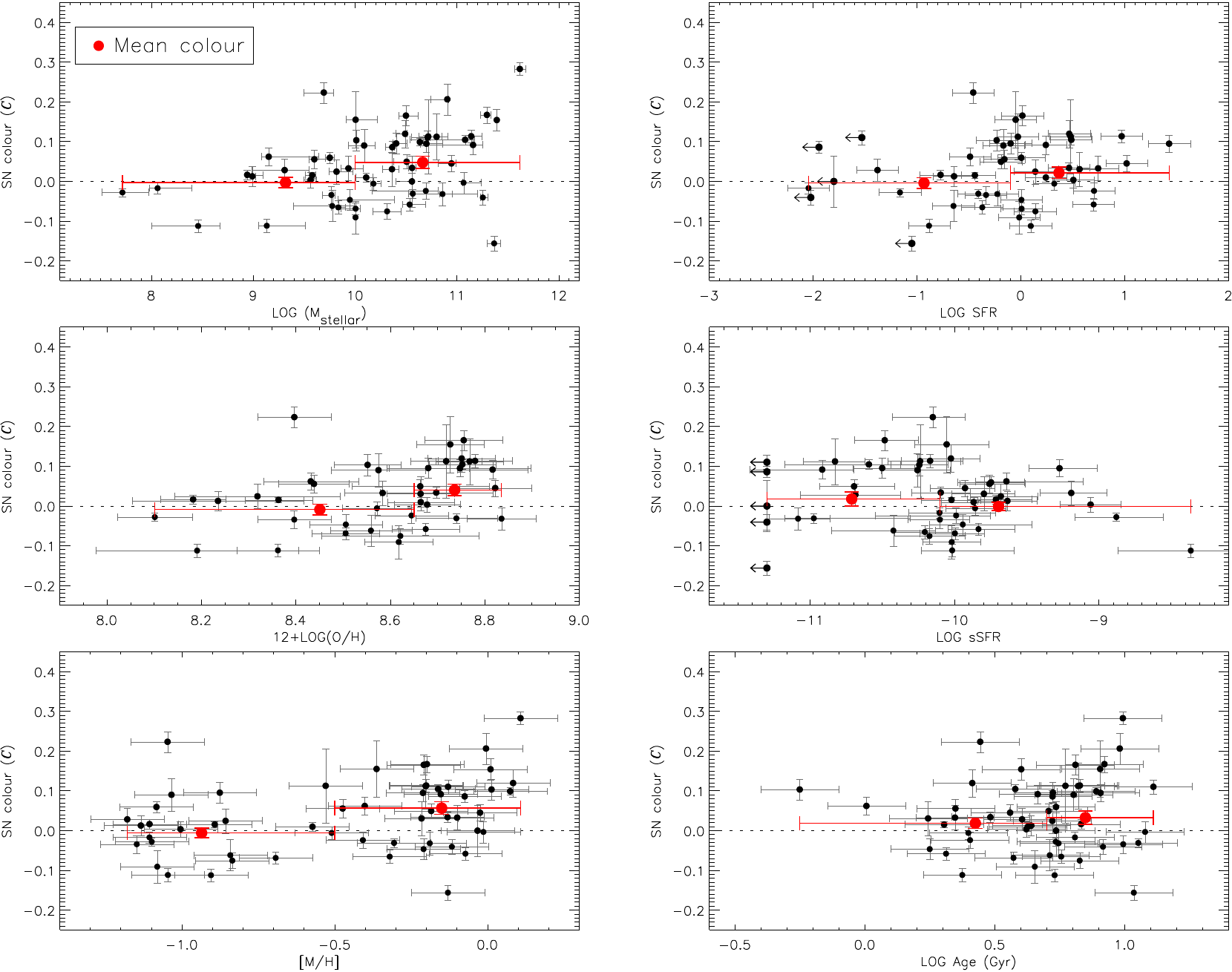}
                \caption{As Fig.~\ref{sn_para1}, but considering the SN
                  colour \col\ in place of stretch.}
        \label{sn_para2}
\end{figure*}

\section{The dependence of SN Ia properties on their host galaxies}
\label{sec:depend-sn-prop}

Having measured various physical parameters of the PTF SN Ia host
galaxies from their spectra and broad-band photometry, we now compare
these parameters with the photometric properties of the SNe. We take
each of the three key SN Ia properties in turn -- stretch (light curve
width), optical colour, and luminosity -- and compare with the
\mstellar, the gas-phase metallicity $12+\log(\rmn{O}/\rmn{H})$, the
stellar metallicity M/H, the stellar age, and the specific SFR (sSFR), the SFR per unit
\mstellar\ \citep[][]{1997ApJ...489..559G}. Compared to the SFR, the
sSFR is a more appropriate indicator to measure the relative
star-formation activity of a galaxy as it measures the star-formation
relative to the underlying galaxy stellar mass.

In this section, we will assess the significance of various relations
between the SN properties and their host galaxies. In each case we
split the sample into two groups. A value of $\mathrm{\log M=10.0}$ is
used to split between the high- and low-\mstellar\ sample, and
$12+\log(\rmn{O}/\rmn{H})=8.65$ and $\mathrm{[M/H]=-0.5}$ are used
(based on the mass-metallicity relation in Fig.~\ref{m-gz_k01} and
relation between gas-phase and stellar metallicities in
Fig.~\ref{sz-gz}) to split between high- and low-metallicity hosts.
For the stellar age, SFR and sSFR, the split points were selected to
make approximately equally sized sub-groups (e.g.,
\citet{2010MNRAS.406..782S}).  The weighted-mean of the residuals in
each group are calculated. The error of the weighted-mean was
corrected to ensure a $\chi^{2}_{\mathrm{red}}=1$.  The linear fitting
is performed by using the Monte Carlo Markov Chain (MCMC) method
\textsc{linmix} \citep{2007ApJ...665.1489K}. To examine the
correlation of the relations, both the Pearson and Kendall correlation
coefficients are also calculated.

\begin{table*}
\centering
\caption{The trend of SN stretch/colour with host parameters.}
\begin{tabular}{lccccc}
\hline\hline
   &        & \multicolumn{2}{c}{SN stretch ($s$)} & \multicolumn{2}{c}{SN colour (\col)} \\
   & Split point  &          N$_{SN}$ & bin difference       &                N$_{SN}$ & bin difference \\
\hline
$\log$ M        & 10.0  & 68 & 0.08 (4.3$\sigma$) & 55 & 0.05 (2.5$\sigma$) \\
12+$\log$(O/H)  & 8.65  & 50 & 0.07 (2.5$\sigma$) & 40 & 0.05 (2.5$\sigma$) \\
$\mathrm{[M/H]}$& -0.5  & 67 & 0.07 (2.5$\sigma$) & 55 & 0.06 (3.1$\sigma$) \\
$\log$ Age      &  0.7  & 67 & 0.08 (3.2$\sigma$) & 55 & 0.01 (0.7$\sigma$) \\
$\log$ SFR      &$-0.1$ & 65 & 0.03 (1.4$\sigma$) & 52 & 0.03 (1.3$\sigma$) \\
$\log$ sSFR     &$-10.1$& 65 & 0.08 (3.1$\sigma$) & 52 & 0.03 (1.2$\sigma$) \\
\hline
\end{tabular}
\label{trend1}
\end{table*}

\subsection{SN Ia stretch}
\label{sec:sn-stretch}
The stretch of a SN Ia is a direct measurement of its light curve
width, a key parameter in the calibration of SNe Ia as distance
estimators \citep{1993ApJ...413L.105P} -- brighter SNe Ia have slower
light curves (a broader width or higher stretch) than their fainter
counterparts. In this study, we restrict our analysis to SNe Ia with
$0.7<s<1.3$, typical of SNe Ia that are used in cosmological analyses
\citep{2011ApJS..192....1C}. This removes one high-stretch and five low-stretch
(sub-luminous) SNe, including the peculiar event PTF09dav
\citep{2011ApJ...732..118S}.

The SN stretch as a function of the host parameters can be found in
Fig.~\ref{sn_para1}. The trend calculated for each case is listed in
Table~\ref{trend1}. When comparing with \mstellar, we recover the
trend seen by earlier studies that lower stretch ($s<1$) SNe Ia are
more likely to be found in massive galaxies than higher stretch
($s\geq1$) SNe Ia
\citep{2009ApJ...691..661H,2009ApJ...707.1449N,2010MNRAS.406..782S}.
Bearing in mind that gas-phase and stellar metallicity strongly
correlates with \mstellar\
\citep[e.g.][]{2004ApJ...613..898T,2005MNRAS.362...41G}, a similar
trend is expected between stretch and metallicity, which is both
observed here (Fig.~\ref{sn_para1}) and has previously been described
in the literature: low-metallicity galaxies preferentially host
brighter SNe Ia (before light curve shape correction). The data also
show that higher-stretch SNe Ia preferentially explode in younger
galaxies, to the extent that there
are few low stretch SNe in hosts with mass-weighted mean ages of less
than $\sim4$\,Gyr. This is consistent with the recent study of
\citet{2013arXiv1309.1182R}, who found that the relation between SN stretch and \mstellar\
is primarily driven by age, as measured by local SFR. We also see moderate trend with sSFR
that galaxies with higher sSFR tend to host brighter SNe Ia. No significant
correlation is found with SFR.

\begin{figure*}
		\includegraphics*[scale=0.7]{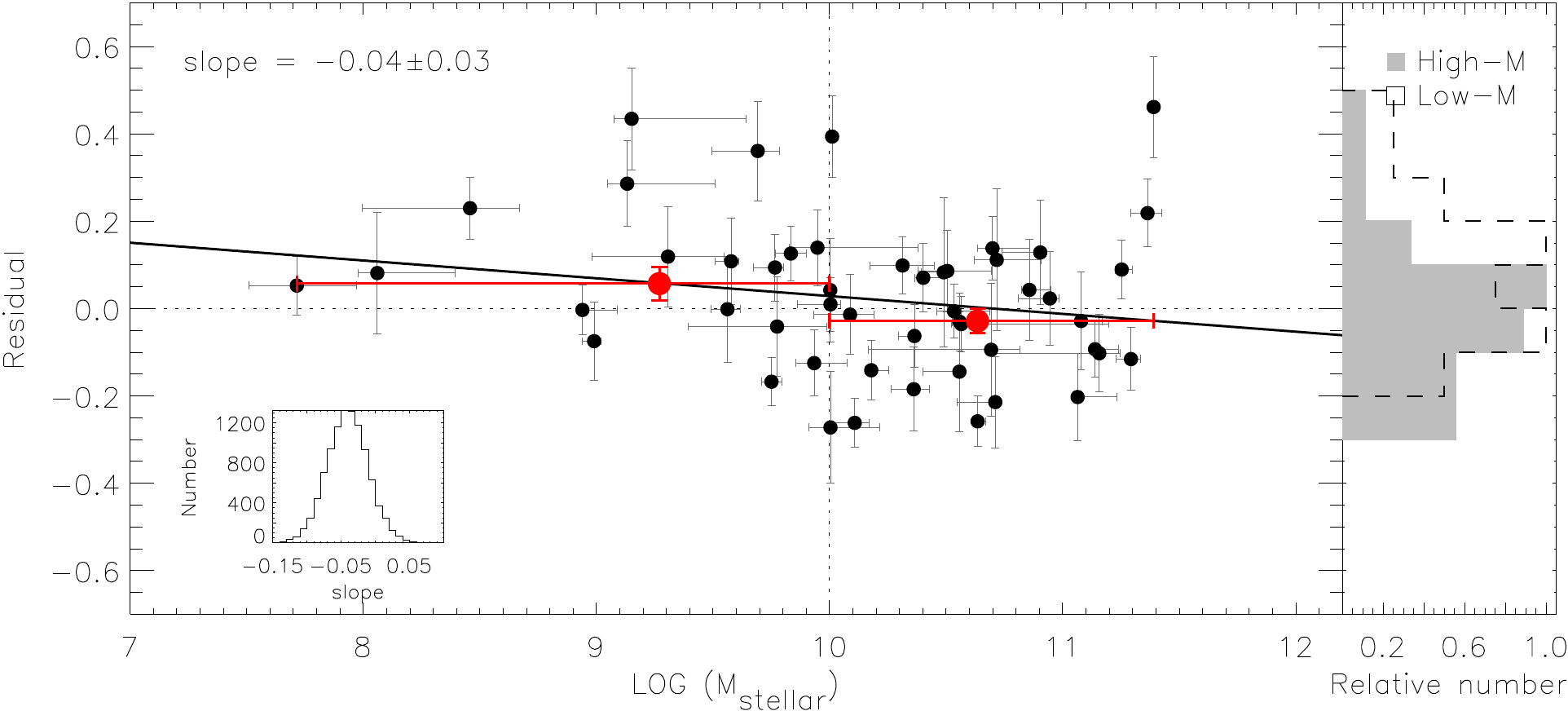}
                \caption{Hubble residuals as a function of host galaxy
                  \mstellar. The vertical dashed line represents the
                  criterion used to split our sample into
                  high-\mstellar\ and low-\mstellar\ galaxies. The red filled
                  circles represent the weighted-mean of the residuals
                  in bins of \mstellar, and their error bars are the
                  width of the bins and the error of the weighted
                  mean.  The histogram on the right shows the
                  distribution of residuals in high-\mstellar\ (filled
                  histogram) and low-\mstellar\ (open histogram). The
                  distribution of the slopes best fit to the data from
                  10,000 MCMC realisations was showed in the sub-plot. The
                  solid line represents the mean slope determined from the distribution.}
        \label{mass-res}
\end{figure*}

\begin{figure*}
		\includegraphics*[scale=0.7]{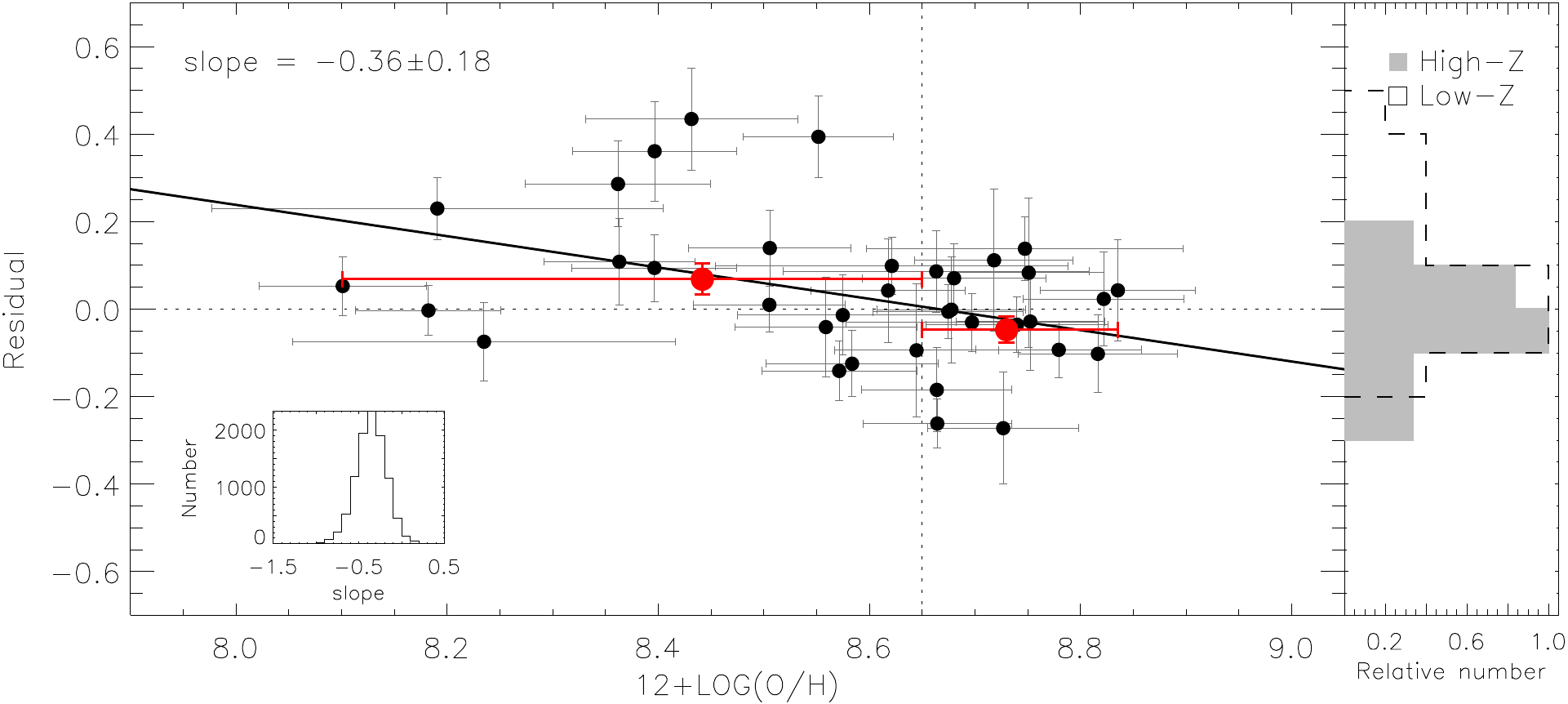}
	\caption{As Fig. \ref{mass-res}, but considering gas-phase metallicity instead of \mstellar.}
        \label{gmetal-res}
\end{figure*}

\subsection{SN Ia colour} 
\label{sec:sn-colour}
We next examine trends with SN Ia colour (\col;
Section~\ref{sec:sn-photometry-light}), shown in Fig.~\ref{sn_para2}.
As for the stretch comparisons, we restrict the SNe to a typical
colour range used in cosmological studies ($\col<0.4$). This removes
five red SNe Ia from our sample.

As star-forming galaxies are expected to contain more dust than
passive galaxies, all other things being equal we would expect them to
host redder SNe Ia.  However, Fig.~\ref{sn_para2} does not show this
effect; if anything SNe Ia in high-sSFR galaxies appear bluer
($\col<0$) than those in low-sSFR galaxies. This may imply an
intrinsic variation of SN colour with host environment that is greater
than any reddening effect from dust. The SN colour as a function of
\mstellar\ does show a trend, with SNe Ia in more massive galaxies
being redder, and both gas-phase and stellar metallicities also show
correlations with SN colour with SNe Ia tending to be redder in
galaxies of higher metallicity.  We will discuss these various trends
involving SN colour in Section~\ref{sec:discussion}.

\begin{figure*}
		\includegraphics*[scale=0.7]{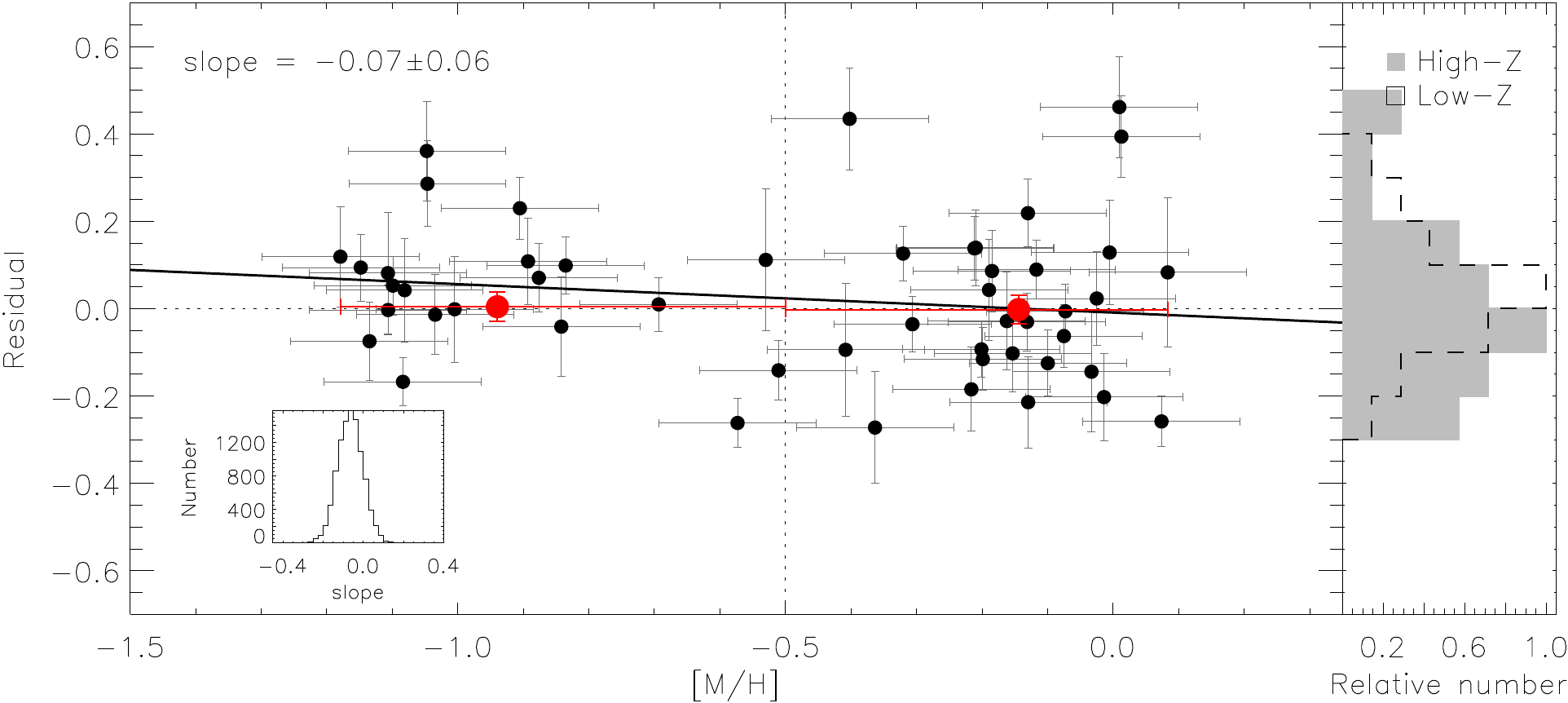}
	\caption{As Fig. \ref{mass-res}, but considering stellar metallicity instead of \mstellar.}
        \label{smetal-res}
\end{figure*}

\begin{figure*}
		\includegraphics*[scale=0.7]{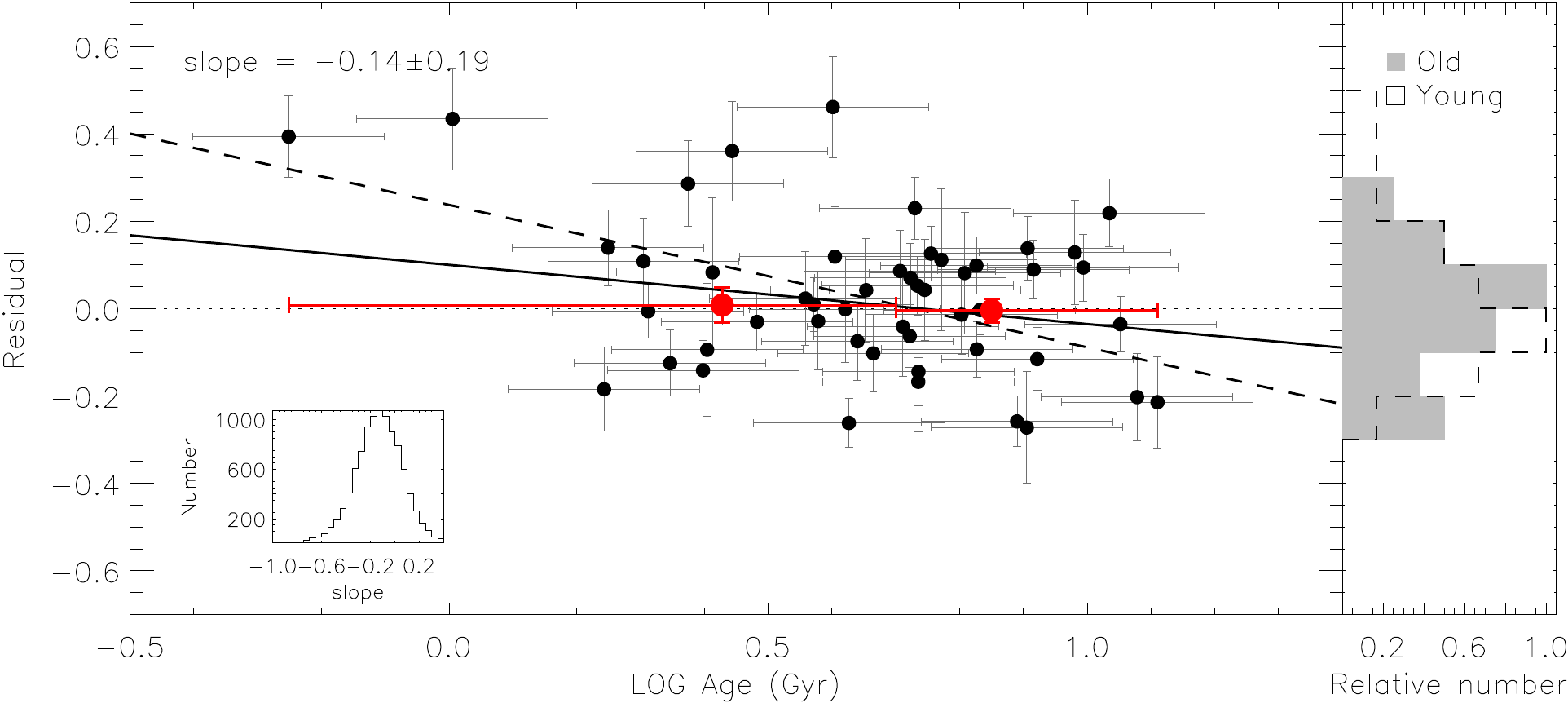}
                \caption{As Fig. \ref{mass-res}, but considering
                  stellar age instead of \mstellar. Here the dashed
                  line includes the two SNe in the youngest host
                  galaxies, which cause a significant trend with
                  Hubble residual, and which otherwise is not
                  present.}
        \label{age-res}
\end{figure*}

\subsection{SN luminosity}
\label{sec:sn-luminosity}
\begin{table*}
\centering
\caption{The trend of Hubble residual with host parameters.}
\begin{tabular}{lccccccc}
\hline\hline
   & Split point & N$_{SN}$ & bin difference & linear trend & probability of & \multicolumn{2}{c}{correlation} \\
   &             &          &    (mag)   &              & negative slope & Pearson & Kendall \\ 
\hline
$\log$ M        & 10.0  & 50 & 0.085 (1.8$\sigma$) & $-0.041 \pm 0.030$ & 91.3\% & $-0.19$ & $-0.13$ \\
12+$\log$(O/H)  & 8.65  & 36 & 0.115 (2.5$\sigma$) & $-0.358 \pm 0.176$ & 98.1\% & $-0.36$ & $-0.22$ \\
$\mathrm{[M/H]}$& -0.5  & 50 & 0.006 (0.1$\sigma$) & $-0.065 \pm 0.063$ & 85.1\% & $-0.13$ & $-0.09$ \\
$\log$ Age      &  0.7  & 48 & 0.012 (0.3$\sigma$) & $-0.135 \pm 0.193$ & 75.7\% & $-0.14$ & $-0.04$ \\
$\log$ sSFR     &$-10.1$& 48 & 0.070 (1.7$\sigma$) & $-0.019 \pm 0.077$ & 58.7\% & 0.04    & $-0.05$ \\
\hline
\end{tabular}
\label{trend2}
\end{table*}

Finally, we turn to the SN luminosity, which we parameterise by
calculating the Hubble residual. This is defined as the difference
between \mbcorr, the observed rest-frame $B$-band SN apparent
magnitude (\mB; Section~\ref{sec:sn-photometry-light}) corrected for
stretch and colour, and \mbmodel, the peak SN magnitude expected in
our assumed cosmological model. At a fixed redshift, a brighter SN Ia
therefore gives a negative Hubble residual. \mbcorr\ is given by
\begin{equation}
\label{eq:mbcorr}
\mbcorr=\mB+\alpha\times(s-1)-\beta\times \col
\end{equation}
and \mbmodel\ by
\begin{equation}
\label{eq:mbmodel}
\mbmodel=5\log_{10}{\mathcal D_L}\left(z;\omatter\right) + \scriptm,
\end{equation}
where $z$ refers to the cosmological redshift in the CMB frame,
${\mathcal D_L}$ is the $c/H_0$ reduced luminosity distance with the
$c/H_0$ factor (here $c$ is the speed of light) absorbed into
\scriptm, the absolute luminosity of a $s=1$ $\col=0$ SN Ia
(eqn.~(\ref{eq:mbcorr})).  Explicitly,
$\scriptm=\absm+5\log_{10}(c/H_0)+25$, where \absm\ is the absolute
magnitude of a SN Ia in the $B$-band. $\alpha$ and $\beta$ are
`nuisance variables' derived from the cosmological fits. In this work
$\alpha=1.45\pm0.12$ and $\beta=3.21\pm0.15$ were obtained. We do not
add any intrinsic dispersion into the Hubble residual uncertainties as
we are, in part, searching for variables which could generate this
extra scatter. However, we are aware of the potential bias that this
could introduce into the nuisance parameters (e.g. $\alpha$ and
$\beta$) when comparing the results to the studies including the
intrinsic dispersion, and thus we caution such a comparison is not
valid. To ensure our SNe are located in the smooth Hubble flow, we
exclude SNe Ia with $z<0.015$, removing one SN from our sample.
The comparisons with the host galaxy parameters can be found in
Figs.~\ref{mass-res} to \ref{age-res}. The trends calculated for Hubble residuals
with host parameters are listed in Table~\ref{trend2}.

The Hubble residuals as a function of host \mstellar\ are shown in
Fig.~\ref{mass-res}. We see only a weak trend that is consistent with
that expected based on earlier work, in the sense that more massive
galaxies preferentially host brighter SNe Ia after stretch and colour
corrections. However, the trend in our data taken in isolation is not
significant: the Hubble residuals of low-\mstellar\ and
high-\mstellar\ bins have a weighted average of $0.057\pm0.038$\,mag
and $-0.028\pm0.028$\,mag respectively, a difference of
$0.085\pm0.047$\,mag. There is a $\sim91\%$ probability the
slope is negative based on 10,000 MCMC realisations.

Figs.~\ref{gmetal-res} and ~\ref{smetal-res} show the Hubble residuals
as a function of gas-phase and stellar metallicity respectively. A
trend with gas-phase metallicity can be seen in the same sense as with
stellar mass: higher-metallicity galaxies tend to host brighter SNe Ia
after stretch and colour corrections. The differences are more
significant than with \mstellar: the Hubble residuals of the
high-metallicity and low-metallicity bins have a weighted average of
$-0.047\pm0.030$\,mag and $0.068\pm0.035$\,mag, respectively, a
difference of $0.115\pm0.046$\,mag.  Fitting a straight line using the
\textsc{linmix} method gives a $\sim98\%$ probability the slope is negative.
The correlation coefficient is $\sim2$ times larger than the
relation between Hubble residuals and \mstellar.  We see no trend with
stellar metallicity.

The Hubble residuals as function of host stellar age are shown in
Fig.~\ref{age-res}; no significant trends are seen. We also see no
trend with sSFR.

\subsection{Comparison to previous studies}
\label{sec:compare-pre-studies}
Many of the relations in the previous section have been studied by
different authors using independent samples of SNe Ia. Since
\mstellar\ is the most straight forward variable to measure, requiring
only broad-band imaging, the most common comparison has been between
Hubble residual and \mstellar, which has been examined with a variety
of samples over a large redshift range
\citep{2010ApJ...715..743K,2010MNRAS.406..782S,2010ApJ...722..566L,2011ApJ...740...92G,2012arXiv1211.1386J,2013arXiv1304.4720C}.
These studies all find that more massive galaxies host brighter SNe
after corrections for light curve shape and colour have been made. The
size of the difference is usually around 0.1\,mag, with a transition
mass of around 10$^{10}$\,$M_{\odot}$. Our result is consistent with
these earlier studies, although at a reduced significance due to a
smaller dataset.

However, our primary goal is to study the metallicity of the SN host
galaxies rather than just \mstellar. Although some studies convert
\mstellar\ into metallicity using average mass-metallicity relations
\citep[e.g.][]{2010MNRAS.406..782S}, it is obviously more useful to
measure the metallicity directly. The first study of this kind was
\citet{2005ApJ...634..210G} who compared the Hubble residuals for 16
local SNe Ia with the gas-phase metallicity of their hosts; no
significant trends were seen.  \citet{2011ApJ...743..172D} studied a
larger sample of $\sim$40 SNe Ia host galaxies from the SDSS-II SN
survey, finding SNe Ia in high-metallicity galaxies to be
$\sim0.1$\,mag ($\simeq4.9\sigma$) brighter than those in
low-metallicity galaxies after corrections, consistent with the
results based on \mstellar. Using the low-redshift SNe studied by
the SNfactory, \citet{2013arXiv1304.4720C} derived the
gas-phase metallicity from 69 SNe Ia hosts and found a difference
$\sim0.1$\,mag ($\simeq2.9\sigma$) difference between high-metallicity
and low-metallicity hosts.  Our results are in good agreement (a
0.115\,mag difference between high- and low-metallicity hosts).
Compared to a metallicity simply converted from host \mstellar, the SN
Ia luminosity shows a stronger dependence on the metallicity derived
via direct emission-line measurements.

For stellar metallicity studies, \citet{2008ApJ...685..752G} studied
29 early-type SN Ia host galaxies by measuring the Lick indices from the SN Ia
host spectra. They found the host stellar metallicity correlates with the
Hubble residual at $\simeq$98 per cent confidence level, although
this technique is not directly comparable to ours.
\citet{2012arXiv1211.1386J} also derived the host stellar metallicity by
measuring the absorption line indices from the SN Ia host spectra, but
did not find a significant trend. We also find no significant trend in our
data.

\citet{2011ApJ...740...92G} determined the mass-weighted average age
of 206 SNe Ia host galaxies by fitting their broad-band photometry.
They found a weak correlation between the Hubble residuals and host age at
$\sim1.9\sigma$.  \citet{2012arXiv1211.1386J} measured the
light-weighted age for the SNe Ia host galaxies using the absorption
line indices but found no significant trend; again we find no trend in
our data.

\citet{2010MNRAS.406..782S} measured photometric-based sSFRs and found
that SNe Ia in low-sSFR hosts appear brighter than those in high-sSFR
hosts at $\simeq2.6\sigma$ significance after $s$ and \col\
corrections.  Similar trends have also been found using
spectroscopy-based sSFRs
\citep{2011ApJ...743..172D,2013arXiv1304.4720C}.  We do not see these
trends in our dataset although this may be due to the relatively small
sample size.

Finally, \citet{2013arXiv1309.1182R} have recently shown that at
least some of the trends with host \mstellar\ may be driven by SNe in
locally passive environments: SNe in massive galaxies with locally passive
environments are systematically brighter than those in locally star forming
environments by $\sim 0.09$\, mag. At the time of writing, it is unclear
whether this trend is due to age or metallicity.

 \begin{figure*}
		\includegraphics*[scale=1]{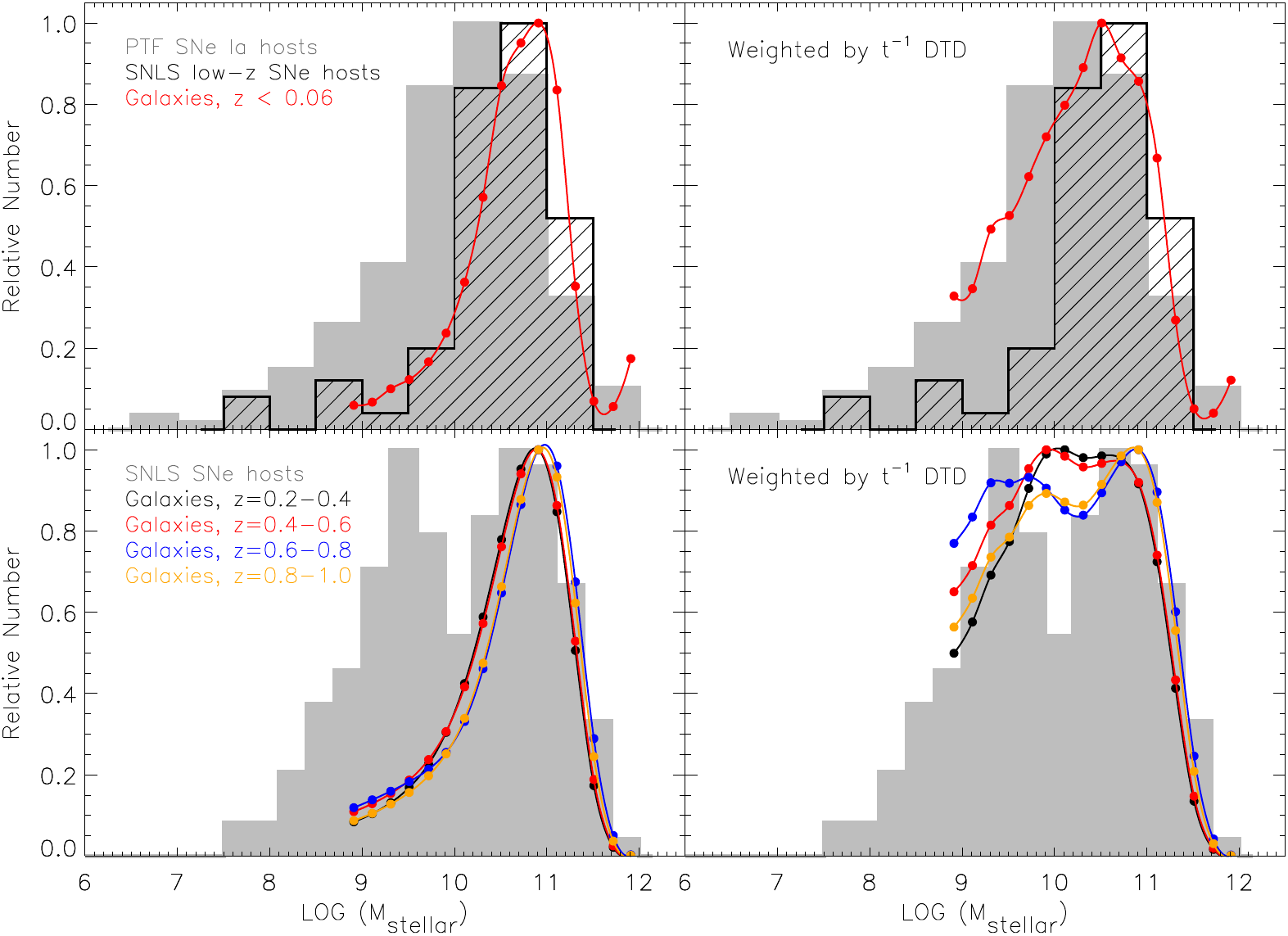}
                \caption{Upper left: The stellar mass distribution of
                  PTF SN Ia host galaxies (grey histogram) compared to
                  the low-redshift SN Ia host sample of
                  \citet{2011ApJS..192....1C} (black histogram). The
                  red solid line with filled circles is the predicted
                  distribution based on the stellar mass function of
                  local galaxies
                  \citep[$z<0.06$;][]{2012MNRAS.421..621B}.  Upper
                  right: The same plot as the upper left panel, but
                  with the predicted SN Ia stellar mass distribution
                  produced by weighting each bin by a $t^{-1}$ DTD.
                  Lower left: The stellar mass distribution of the SNe
                  Ia host galaxies in the SNLS sample of
                  \citet{2010MNRAS.406..782S} (grey histogram). The
                  lines with filled circles in different colours
                  represent the mass contributions derived from the
                  galaxy stellar mass functions in different redshift
                  ranges studied by \citet{2009ApJ...707.1595D}.
                  Lower right: The same plot as left panel, but with
                  prediction weighted by the $t^{-1}$ DTD.}
        \label{mass_compare}
\end{figure*}

\section{Discussion}
\label{sec:discussion}
\subsection{The host stellar mass distribution}
\label{sec:host-stellar-mass-2}
\begin{figure*}
		\includegraphics*[scale=1]{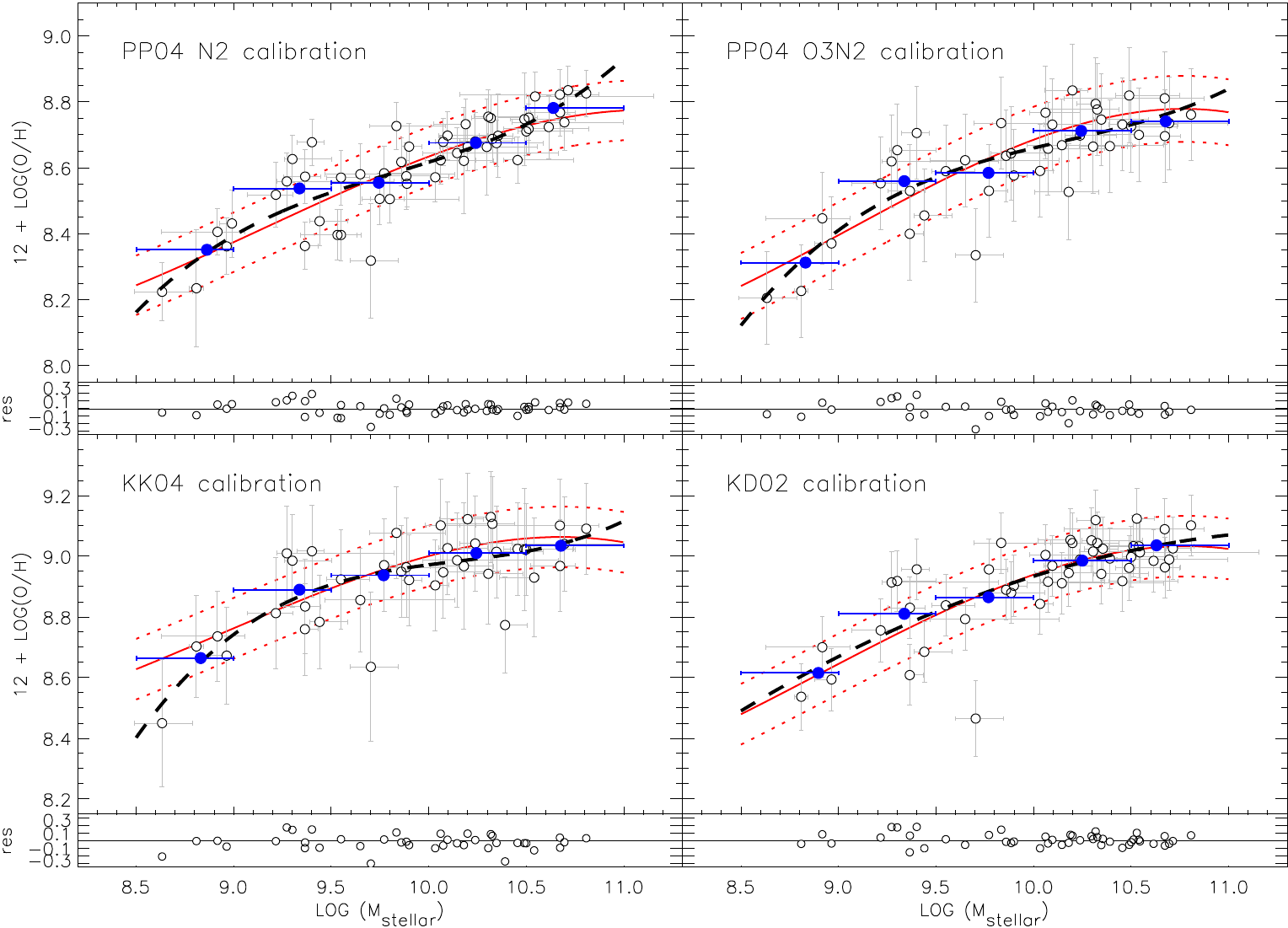}
	\caption{The metallicity-mass relations compared by different
	metallicity calibrations. Upper panels: \citetalias{2004MNRAS.348L..59P} N2 (left)
	and \citetalias{2004MNRAS.348L..59P} O3N2 (right) calibrations.
	Lower panels: KK04 (left) and KD02 (right) calibrations.
	The open circles show the measurements in this work. The blue
	filled-circle is the mean metallicity in bins of \mstellar.
	The best fit to field galaxies given by \citet{2008ApJ...681.1183K}
	with the range of one r.m.s scatter was over-plotted
	(in red line and red dashed-line, respectively). The black
	dashed-line is the best fit to the measurements in this work.
	The sub-plot in the bottom of each panel represents the residuals
	of our measurements from KE08's best fit.}
        \label{m-gz_compare}
\end{figure*}

Unlike galaxy-targeted supernova surveys that are biased towards surveying
brighter and more massive galaxies, the host galaxies in rolling
searches such as PTF should represent a wider range of SN
environments. A key test of this is the host galaxy stellar mass
distribution (Fig.~\ref{sample_selection};
Section~\ref{sec:observ-data-reduct}), the form of which should be a
combination of the underlying galaxy stellar mass function and the
mean SN Ia rate as a function of stellar mass, and which should be
able to be reproduced from a qualitative knowledge of both. We examine
this in this section.

We begin with the galaxy stellar mass function (GSMF) in the local
universe, defined as the number of galaxies per logarithmic bin in
stellar mass. For this study we adopt the GSMF of
\citet{2012MNRAS.421..621B}, the redshift range of which ($z<0.06$) is
similar to this work. In each bin in stellar mass, we divide the GSMF
by the total mass (stellar mass multiplied by the GSMF) in that bin.
We then over-plot the observed SN Ia host galaxy stellar mass
distribution. We compare both to a SN Ia host mass distribution drawn
primarily from galaxy-targeted searches \citep[the low-$z$ sample
compilation of][]{2011ApJS..192....1C}. The
result can be seen in the upper left panel of Fig.~\ref{mass_compare}.

The host stellar mass distribution for the galaxy-targeted SN Ia
sample is similar to that expected based on the GSMF, but is obviously
different to our PTF sample at smaller stellar masses. However, this
test assumes that the SN Ia rate is simply proportional to the stellar
mass of the host. While this may be approximately correct in the more
massive systems, it is known to be incorrect in lower stellar mass
systems which have a larger fraction of younger potential progenitor
systems or higher sSFRs (Fig.~\ref{m-sfr}) -- the SN Ia rate is not
simply proportional to stellar mass
\citep[e.g.][]{2005A&A...433..807M,2006MNRAS.370..773M,2006ApJ...648..868S,2012ApJ...755...61S}.
In practise, a delay-time distribution (DTD; the distribution of times
between the progenitor star formation and the subsequent SN Ia
explosion) with the SN Ia rate proportional to $t^{-1}$ is favoured
by most recent data \citep[e.g.][]{2012PASA...29..447M}. If we assume
this DTD, and the relation between \mstellar\ and galaxy age
determined by \citet{2005MNRAS.362...41G}, a revised distribution of
SN Ia host galaxy stellar masses can be formed by weighting each mass
bin by a $t^{-1}$ DTD.

The results are shown in Fig.~\ref{mass_compare} (upper right): the
effect of the $t^{-1}$ DTD is to increase the contribution from SNe in
less massive (younger) galaxies. A $\chi^2$ is calculated between
galaxy stellar mass distribution and SN Ia host galaxy stellar mass distribution.
The $\chi^2$ drops from 123.18 to 12.21 after considering a $t^{-1}$ DTD to the galaxy
stellar mass distribution. Indeed, assuming a $t^{-1}$ DTD and a
simple scaling between stellar mass and age allows an excellent
reproduction of the observed host galaxy stellar mass distribution.

A similar comparison can be made to the stellar mass distribution from
the SNLS sample \citep{2010MNRAS.406..782S}. As seen in the lower
panels of Fig.~\ref{mass_compare}, the SNLS sample contains more lower
stellar-mass galaxies than our PTF sample even though the selection of
SNe should be similar. Using the GSMF of \citet{2009ApJ...707.1595D}
over $0.2<z<1.0$, and the same technique as above, again a good
agreement between the observed host galaxy \mstellar\ distribution and
that derived from the GSMF is achieved. Thus the difference in the
stellar mass distributions of the PTF and SNLS host galaxies can be
explained by evolution in the field galaxy population from which the
hosts are drawn where there is an excess in low-mass galaxies for the stellar
mass distributions at high redshifts.

\subsection{The mass--metallicity relation of SN Ia host galaxies}
\label{sec:m-z-relation}
\begin{figure*}
		\includegraphics*[scale=0.9]{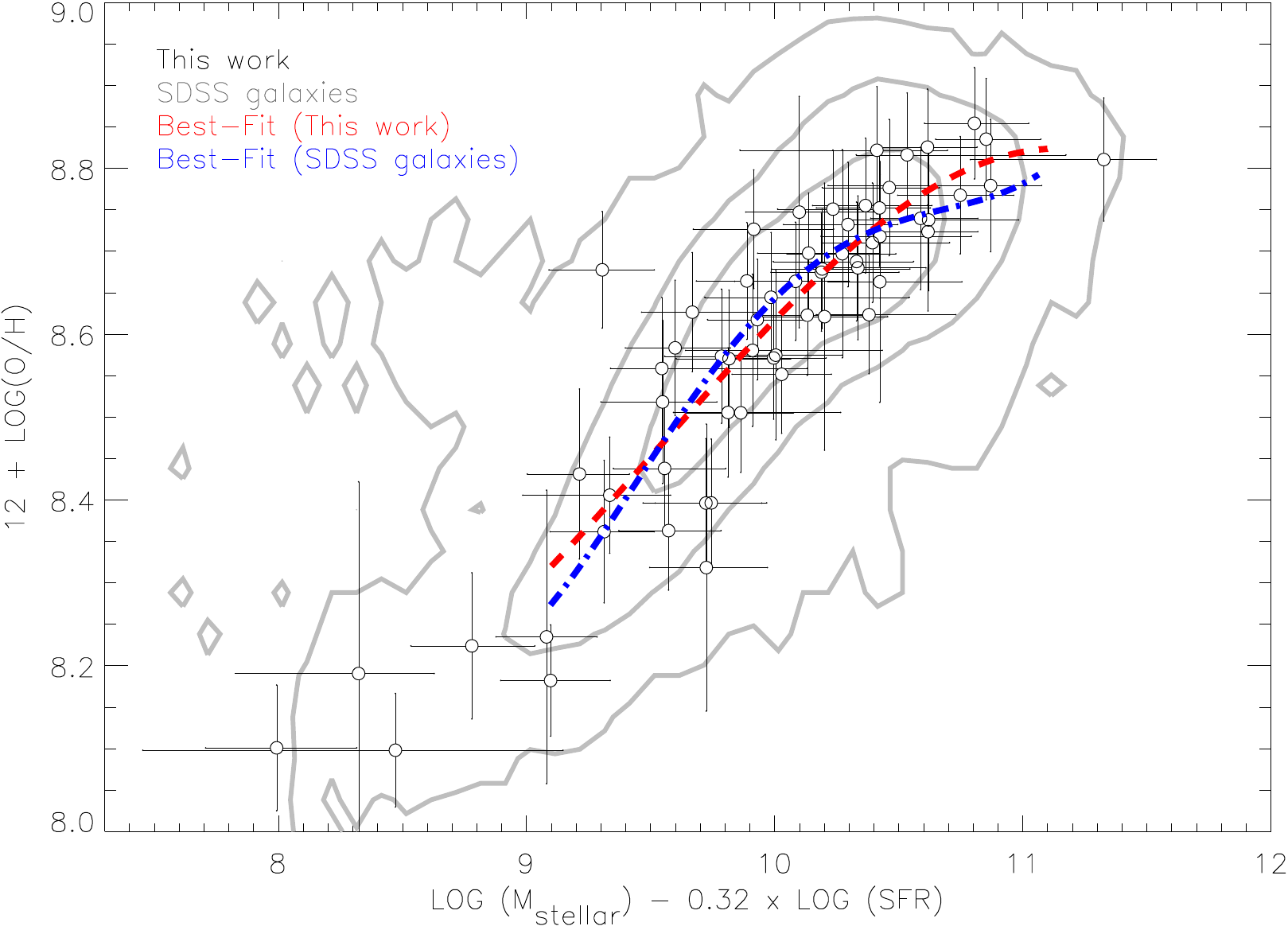}
                \caption{The fundamental metallicity relation (FMR)
                  derived from our data. The open circles
                  show the measurement of SN hosts. The red dashed
                  line is the best-fit to it. The grey contours show the
                  sample including $68\%$, $95\%$ and $99.7\%$ of SDSS galaxies from
                  \citet{2010MNRAS.408.2115M}. The blue dot-dashed line is the best
                  fit to the SDSS galaxies.}
        \label{FMR}
\end{figure*}

\begin{figure*}
		\includegraphics*[scale=0.9]{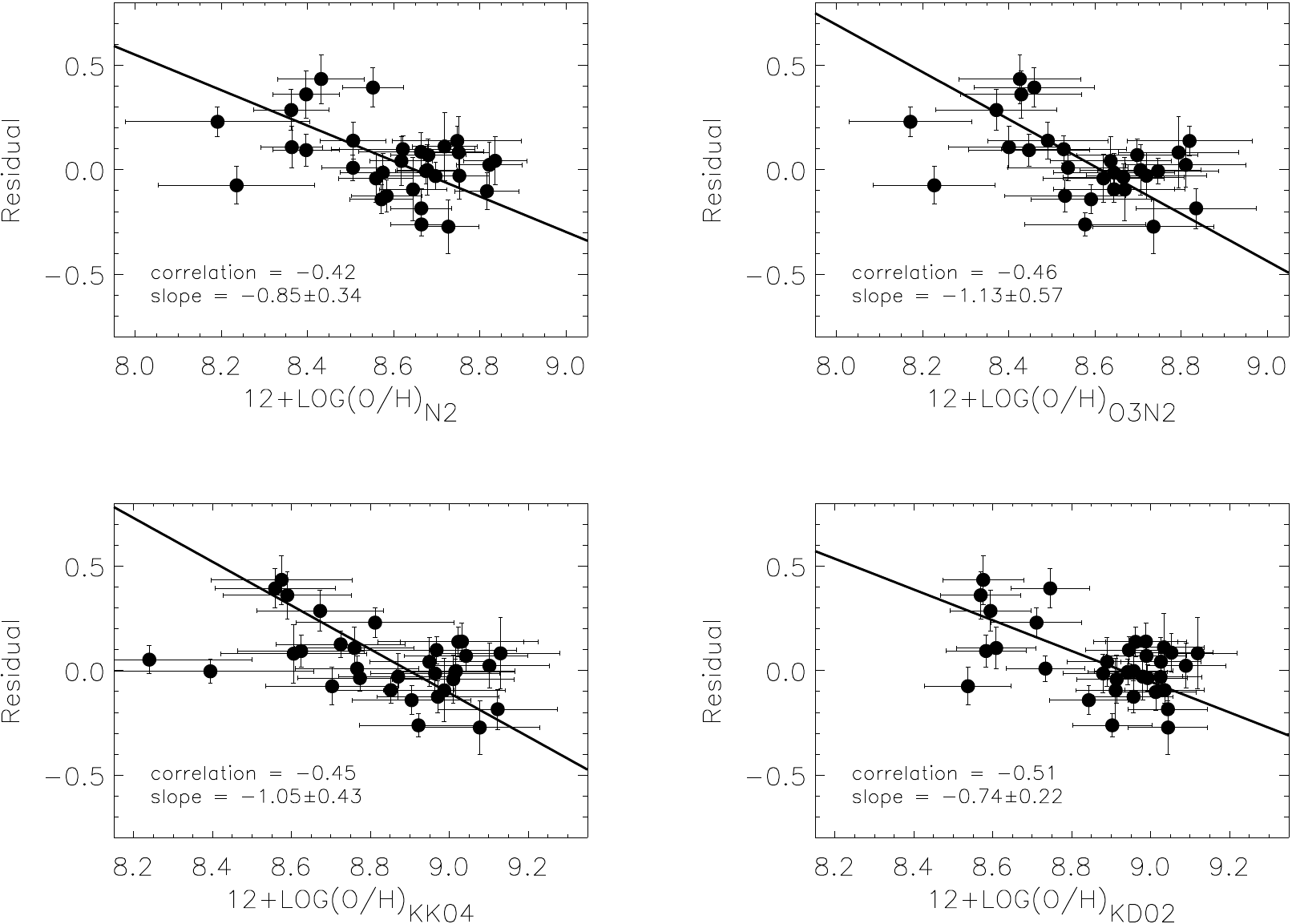}
                \caption{The Hubble residuals as function of host
                  galaxy gas-phase metallicity using four different
                  calibrations. The best linear fit is shown
                  in solid line in each panel.}
        \label{res_compare}
\end{figure*}

As well as impacting on the observed photometric properties of SNe Ia,
the metallicity of the progenitor star may also impact on the SN Ia
rate. An increased rate with lower metallicity may be expected as
stars with a lower metallicity generally form more massive white
dwarfs and therefore may more easily approach the Chandrasekhar mass
limit \citep{2011arXiv1106.3115K}. A decreased rate with lower
metallicity may be expected in some single degenerate scenarios as the
lower metallicity inhibits accretion onto the white dwarf due to lower
opacities in the wind \citep{1998ApJ...503L.155K,2009ApJ...707.1466K}.
Observationally, there is some evidence that prompt SNe Ia are more
prevalent (or explode with a brighter luminosity) in metal-poor
systems \citep*{2009ApJ...704..687C}.

Such effects may impact on the observed SN Ia host mass--metallicity
relation; if any of the putative metallicity effects lead to SNe Ia
preferentially occurring in low or high metallicity galaxies, then the
mass--metallicity relation for SN Ia hosts would be offset from that
of field galaxies (e.g., at fixed galaxy stellar mass, the SN Ia host
galaxies may systematically have lower or higher metallicities than
the field galaxies). We compare our mass--metallicity relation with
those derived for field galaxies (Fig.~\ref{m-gz_k01}), and compare
the use of four different gas-phase metallicity calibrations in
Fig.~\ref{m-gz_compare}. We fit these mass--metallicity relations
using the same functional form as described in KE08 at $8.5 <
\mathrm{logM} < 11.0$. Note that although we estimate \mstellar\ by
fitting broad-band photometry instead of using the spectroscopic
indices of KE08, previous work has shown that the two different
approaches provide consistent results
\citep[0.001\,dex;][]{2012ApJ...755...61S}.  Thus the different
\mstellar\ determination techniques should have a negligible effect on
our results.

The SN Ia host mass--metallicity relations are consistent with the
fits from KE08, with weighted mean offset $\sim0.01$\,dex
($\lesssim1\sigma$ significance). This is
consistent with \citet{2013arXiv1304.4719C}, who found that SN Ia
hosts in their sample also show a good agreement with field galaxy
mass-metallicity relation.

However, this comparison has one potential systematic -- at fixed
\mstellar, other variables may affect the SN Ia rate, for example the
number of young stars (or the SFR). Indeed,
\citet{2010MNRAS.408.2115M} showed that the observed mass--metallicity
relation could be a projection of a more general relation between
\mstellar, gas-phase metallicity, and SFR, which together can be
described using a fundamental metallicity relation (FMR). The FMR can
be defined as the relation between gas-phase metallicity and
$\mathrm{log(\mstellar)-\alpha\times log(SFR)}$, where $\alpha$ is a
parameter determined to minimize the scatter of the metallicities.
\citet{2010MNRAS.408.2115M} found $\alpha=0.32$ produced the
minimum dispersion in metallicity.

We therefore construct the FMR for the 61 SN Ia hosts with a measure
of \mstellar, SFR and gas-phase metallicity. For comparison, we also
determined the FMR for SDSS field galaxies using the same parent
sample as \citet{2010MNRAS.408.2115M}. Similar quality cuts as
described in \citet{2010MNRAS.408.2115M} have been performed. In
addition, we applied aperture corrections to the SDSS sample, and
select galaxies within a similar redshift range as our host sample
(median $z\sim0.07$).  The results are shown in Fig.~\ref{FMR}. We
found no significant difference between the FMR for SN Ia hosts and
that of the SDSS galaxies. The weighted mean offset between the
SNe Ia hosts and best-fit from SDSS galaxies is $0.005\pm0.011$\,dex.

In summary, by examining both the mass--metallicity relation and FMR
of SN Ia host galaxies, we find a good agreement with the same
relations derived from field galaxies, suggesting SN Ia host galaxies
and normal field galaxies follow similar relations. This in turn
suggests that the effect of metallicity on the SN Ia rate must be
small.

\subsection{The effect of metallicity on SN Ia luminosities}
\label{sec:effect-metall-sn}
The peak luminosity of SNe Ia is powered by the radioactive decay of
\Nifs\ synthesised during the explosion. \citet{2003ApJ...590L..83T}
showed that the observed scatter in metallicity could introduce a
$25\%$ variation in the mass of \Nifs\ synthesized by SNe Ia.
Metal-rich stars tend to synthesize more neutron-rich (and stable)
$^{58}$Ni instead of the \Nifs\ that powers the SN Ia luminosity. As a
result, and all other variables being equal, intrinsically fainter SNe
Ia are expected to explode in higher-metallicity environments.

However, \citet{2011MNRAS.414.1592B} showed that SN Ia metallicities
can be reasonably estimated by the host galaxy metallicity, and are
better represented by gas-phase metallicity than by stellar
metallicity. In this study we used the host gas-phase metallicity as a
proxy for the progenitor metallicity to study the metallicity effects
on SNe Ia.  Fig.~\ref{res_compare} shows the dependence of Hubble
residuals on metallicities based on different calibrations. Again, the
relative metallicity conversions were not applied here, therefore the
number of hosts which are available for different metallicity
calibrations could be different.  The results showed a good
consistency between different calibrations. The slopes range from
$-0.74$ to $-1.13$, with the Pearson correlation
coefficients range from $-0.42$ to $-0.51$. This suggests that the
correlation between Hubble residual and gas-phase metallicity is
independent of the calibration methods at least at the level of
precision probed here.

\begin{table}
\centering
\caption{Kendall rank correlation coefficients between Hubble residual (HR), gas-phase metallicity, \mstellar\ and stellar age.}
\begin{tabular}{l|cccc}
\hline\hline
  & HR & 12+$\log$(O/H) & \mstellar & Age \\
\hline
HR & -- & $-0.22 $ & $-0.13$ & $-0.04$ \\
12+$\log$(O/H)  &  & -- & 0.49 & 0.03 \\
\mstellar &  &  & -- & 0.35 \\
Age &  &  &  & -- \\
\hline
\end{tabular}
\label{kendall_correlation}
\end{table}

Followed the procedure described in \citet{1994MNRAS.268..305H},
the Kendall rank correlation coefficients between
Hubble residual, gas-phase metallicity, \mstellar\, and stellar age are listed in
Table~\ref{kendall_correlation}.  Our results show that the SN Ia
luminosity has the strongest dependence on the host gas-phase
metallicity compared to \mstellar\ or stellar age.  We also found that
the correlation coefficient between Hubble residuals and \mstellar\
is similar to the value multiplicatively combining the correlation
coefficients of the Hubble residual--metallicity and
\mstellar\--metallicity relations, from which it can be inferred that
the correlation between Hubble residuals and \mstellar\ may be a
consequence of the Hubble residual--metallicity relation and the
strong correlation between \mstellar\ and metallicity.

\citet{2013ApJ...764..191H} applied the FMR to the SN Ia host galaxies
using broad-band colours alone.  They found the scatter of Hubble
residual is greatly reduced by using the FMR instead of just the
\mstellar, which in turn implies that metallicity may be the
underlying cause of the correlation between Hubble residual and
\mstellar. By directly measuring the gas-phase metallicity of SN Ia
host galaxies we can also show that it has a more significant effect on
SN Ia luminosity than \mstellar\ or stellar age.

\subsection{The SN Ia intrinsic colour}
\label{sec:sn-ia-intrinsic}
\begin{figure*}
		\includegraphics*[scale=0.7]{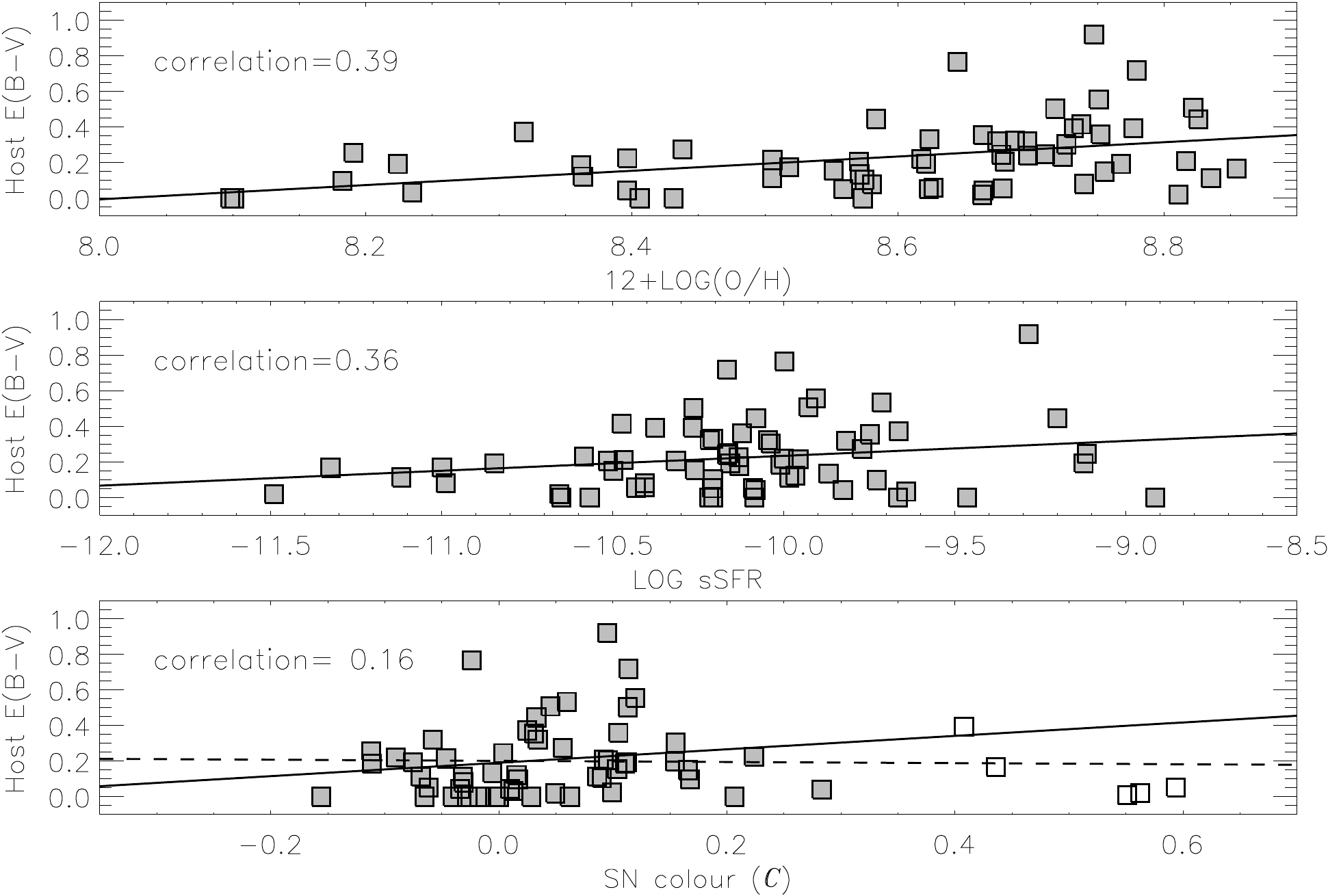}
                \caption{The colour excess of the host galaxies
                  $E(B-V)$ as function of gas-phase metallicity, specific SFR
                  and SN colour \col.
                  The solid line in each panel represents
                  the best linear fit to the data. The SNe with $\col>0.4$
                  are shown in open squares in the lower panel,
                  and the dashed line is the best linear fit
                  to the data including these red SNe.
                  }
        \label{ebv-hostpara}
\end{figure*}

In Section~\ref{sec:sn-colour} we examined the correlations between SN
colour and host parameters, and found that the SNe Ia in
high-metallicity and/or low-sSFR hosts appear to be redder. However,
the SN colour discussed here (SiFTO \col) is not an `intrinsic' SN
colour, as dust extinction from the host galaxy may also contribute to
the observed SN colour variation.  Therefore it is useful to compare
the SN colour measures to independent measures of host extinction to
assess the effect of dust extinction from the host galaxies.

Fig.~\ref{ebv-hostpara} shows the host colour excess $E(B-V)$
determined from the Balmer decrement as function of host parameters
and SN colour.  Mild correlations between $E(B-V)$ and the gas-phase
metallicity/sSFR are found, with Pearson correlation coefficients of
0.39 and 0.36 for gas-phase metallicity and sSFR, respectively.
However, we see no significant correlation with SN colour: the SN
colour appears independent of the host galaxy $E(B-V)$. This
independence become even more evident when including those red SNe Ia
with $\col>0.4$ previously excluded in this study.

There are two possibilities that could cause this. The first is that the 
bulk of the SN Ia colour variation is intrinsic to the SN event. Previous
studies have shown that the SN Ia intrinsic colour could be altered
systematically by changing the metallicity of the progenitor \citep*{1998ApJ...495..617H,2001ApJ...557..279D}.
There is some observational evidence for this: \citet{2013arXiv1304.4720C}
recently showed that SNe Ia in high-metallicity hosts appear redder, and
we also find a similar dependence of SN Ia colour on host gas-phase and
stellar metallicities in Fig.~\ref{sn_para2}.

A second possibility is that there is dust local to the SN explosion that affects the colour,
for example circumstellar dust \citep{2005ApJ...635L..33W,2008ApJ...686L.103G,2011ApJ...735...20A},
but would not be traced by photons emerging from HII regions in the host galaxy. This interpretation
is supported by evidence that the $B-V$ colour at maximum of SNe Ia correlates with the strength
of narrow, blueshifted Na I D features in SN Ia spectra \citep{2013arXiv1308.3899M,2013ApJ...772...19F},
which likely trace the presence of circumstellar material \citep{2007Sci...317..924P,2011Sci...333..856S}.

\section{Conclusions}
\label{sec:conclusions}

In this paper, we have derived the parameters of a sample of 82 SN Ia
host galaxies from the Palomar Transient Factory (PTF), and used it to
examine the relationships between SNe Ia and their hosts.  The host
galaxy parameters have been determined from both photometric and
spectroscopic data. In particular, we have derived star-formation
rates, stellar masses, gas-phase and stellar metallicities, and
stellar ages for the host galaxies. Our main findings are:

\begin{enumerate}
\item[$\bullet$] Previously observed correlations between SN Ia
  properties and their host parameters are recovered in this work. In
  particular, we show that the SN light-curve width (stretch) has a
  strong dependence on host galaxy age and stellar mass -- fainter,
  faster evolving (lower stretch) SNe Ia tend to be hosted by older,
  massive galaxies.
\item[$\bullet$] For the SN Ia colour, we have shown that redder SNe
  Ia have a tendency to explode in more metal-rich galaxies. However,
  we found no relation between SN colour and the colour excess of the
  host galaxies as measured from the Balmer decrement, suggesting that
  the bulk of the SN Ia colour variation is intrinsic and not
  dependent on host galaxy extinction.
\item[$\bullet$] The dependence of the SN Ia Hubble residuals on host
  gas-phase metallicities was also confirmed.  SNe Ia in metal-rich
  galaxies are $\sim0.1$\,mag brighter than those in metal-poor
  galaxies after light-curve shape and colour corrections.  This
  dependence does not depend on different metallicity calibrations.
  The correlation derived between Hubble residual and gas-phase
  metallicities is about two times stronger than for stellar mass.
  That implies that the host galaxy metallicity may be the underlying
  cause of the well-established relation between SN Ia luminosity and
  stellar mass.
\item[$\bullet$] We showed that the stellar mass distribution of the
  PTF and SNLS SN Ia host galaxies are quite different, with SNLS
  possessing many more low-mass host galaxies than PTF. However this
  can be understood and reproduced by a combination of a
  redshift-dependent galaxy stellar mass function (GSMF), and a SN Ia
  rate inversely proportional to the age of the galaxy (a $t^{-1}$
  DTD).
\item[$\bullet$] Finally, we compared the mass-metallicity relation
  for the SN Ia hosts to that of field galaxies drawn from SDSS.  We
  found no significant difference between the two relations, a result
  that is not sensitive to the metallicity calibrations adopted. In
  addition, we derived the fundamental metallicity relation
  \citep[FMR;][]{2010MNRAS.408.2115M} for the SN Ia hosts and also
  found it to be similar to that measured from field galaxies. This
  suggests that metallicity has a negligible effect on the SN Ia rate.
\end{enumerate}

This study has emphasised the important role of the host galaxies of
SNe Ia in influencing the SN explosion properties, with SN Ia
properties showing considerable dependence on their host galaxy
parameters.  From a cosmological perspective, the precision of the
cosmology can be improved by correcting these biases introduced by
host galaxies. It can also shed light on the properties of SN Ia
progenitors. Therefore it is of great importance for future SN Ia
surveys to study both the SNe and the galaxies in which they explode.

\section*{Acknowledgements}

MS acknowledges support from the Royal Society. A.G. acknowledges support
from the EU/FP7 via and ERC grant, funding from the ISF and BSF, and the
Minerva ARCHES and Kimmel awards.

Based on observations obtained at the Gemini Observatory, which is operated by the
Association of Universities for Research in Astronomy, Inc., under a
cooperative agreement with the NSF on behalf of the Gemini
partnership: the National Science Foundation (United States), the
National Research Council (Canada), CONICYT (Chile), the Australian
Research Council (Australia), Minist\'{e}rio da Ci\^{e}ncia,
Tecnologia e Inova\c{c}\~{a}o (Brazil) and Ministerio de Ciencia,
Tecnolog\'{i}a e Innovaci\'{o}n Productiva (Argentina). Based on
Gemini progams GN-2010B-Q-111, GS-2010B-Q-82, GN-2011A-Q-82,
GN-2011B-Q-108, GN-2012A-Q-91, GS-2012A-Q-3, GN-2012B-Q-122, and
GS-2012B-Q-83 for the host galaxy observations,
and GN-2010A-Q-20, GN-2010B-Q-13, GN-2011A-Q-16 and GS-2009B-Q-11
for the SN observations.
The William Herschel Telescope is operated on the
island of La Palma by the Isaac Newton Group in the Spanish
Observatorio del Roque de los Muchachos of the Instituto de
Astrof�sica de Canarias.  Observations obtained with the Samuel
Oschin Telescope at the Palomar Observatory as part of the Palomar
Transient Factory project, a scientific collaboration between the
California Institute of Technology, Columbia University, Las Cumbres
Observatory, the Lawrence Berkeley National Laboratory, the National
Energy Research Scientific Computing Center, the University of Oxford,
and the Weizmann Institute of Science.  Some of the data presented
herein were obtained at the W.M. Keck Observatory, which is operated
as a scientific partnership among the California Institute of
Technology, the University of California and the National Aeronautics
and Space Administration. The Observatory was made possible by the
generous financial support of the W.M. Keck Foundation.
Based on observations collected at the European Organisation for Astronomical
Research in the Southern Hemisphere, Chile, under program IDs
084.A-0149 and 085.A-0777.
Observations obtained with the SuperNova Integral Field Spectrograph
on the University of Hawaii 2.2-m telescope as part of the
Nearby Supernova Factory II project, a scientific collaboration between
the Centre de Recherche Astronomique de Lyon, Institut de Physique Nucl'eaire de Lyon,
Laboratoire de Physique Nucl'eaire et des Hautes Energies,
Lawrence Berkeley National Laboratory, Yale University, University of Bonn,
Max Planck Institute for Astrophysics, Tsinghua Center for Astrophysics,
and Centre de Physique des Particules de Marseille.
This research has made use of the NASA/IPAC Extragalactic Database (NED) which is
operated by the Jet Propulsion Laboratory, California Institute of
Technology, under contract with the National Aeronautics and Space
Administration.

This publication has been made possible by the participation of more
than 10 000 volunteers in the Galaxy Zoo: Supernovae project
(\url{http://supernova.galaxyzoo.org/authors}).

\bibliographystyle{mn2e}
\bibliography{pan}

\begin{table*}
\centering
\caption{The observing log of full sample used in this paper.}
\begin{threeparttable}[b]
\renewcommand{\arraystretch}{0.93}
\scriptsize
\begin{tabular}{lcccccccc}
\hline\hline
SN name & \multicolumn{2}{c}{Coordinate}  & Instrument & \multicolumn{2}{c}{Exp. time\tnote{a}}  & \multicolumn{2}{c}{redshift $(z)$\tnote{b}}  & S/N \\
 & \multicolumn{1}{c}{(R.A.)} & \multicolumn{1}{c}{(Dec.)} & &(Blue)&(Red)& \multicolumn{1}{c}{(Spec.)} & \multicolumn{1}{c}{(CMB)} &  \\  
\hline
PTF09dav   &   22:46:52.95  &   +21:38:21.5  &    Gemini+GMOS   & 3000 & 3000 & 0.0371 & 0.0359 &    42\\
PTF09dlc   &   21:46:30.02  &   +06:25:07.3  &    Gemini+GMOS   & 4000 & 4000 & 0.0672 & 0.0660 &    15\\
PTF09dnl   &   17:23:41.69  &   +30:29:51.0  &    Gemini+GMOS   & 3000 & 3000 & 0.0237 & 0.0236 &    13\\
PTF09dnp   &   15:19:25.34  &   +49:30:05.0  &    Gemini+GMOS   & 3600 & 3600 & 0.0370 & 0.0373 &    28\\
PTF09dxo   &   02:01:47.64  &   -07:05:34.4  &    Gemini+GMOS   & 4000 & 3000 & 0.0519 & 0.0510 &    35\\
PTF09dxw   &   23:44:50.66  &   +27:37:28.9  &    Gemini+GMOS   & 3000 & 3000 & 0.0293 & 0.0282 &    46\\
PTF09dza   &   21:03:31.68  &   -04:20:35.5  &    Gemini+GMOS   & 4000 & 4000 & 0.0832 & 0.0822 &    44\\
PTF09edr   &   00:11:42.89  &   +30:14:22.6  &    Gemini+GMOS   & 3000 & 3000 & 0.0857 & 0.0846 &    35\\
PTF09eoi   &   23:24:12.96  &   +12:46:46.6  &    Gemini+GMOS   & 3000 & 3000 & 0.0410 & 0.0398 &    39\\
PTF09fox   &   23:20:47.79  &   +32:30:06.1  &    Gemini+GMOS   & 3000 & 3000 & 0.0718 & 0.0707 &    33\\
PTF09foz   &   00:42:11.90  &   -09:52:54.8  &    Gemini+GMOS   & 3000 & 3600 & 0.0543 & 0.0532 &    35\\
PTF09gce   &   23:38:49.63  &   +28:16:13.8  &    Gemini+GMOS   & 3000 & 3000 & 0.0575 & 0.0564 &    30\\
PTF09gon   &   22:42:45.53  &   +05:09:28.8  &    Gemini+GMOS   & 4000 & 3600 & 0.0680 & 0.0668 &    36\\
PTF09gul   &   01:28:23.90  &   +33:51:40.7  &    Gemini+GMOS   & 3600 & 3600 & 0.0725 & 0.0716 &    39\\
PTF09hpl   &   01:21:29.18  &   +00:06:28.8  &    Gemini+GMOS   & 4000 & 3600 & 0.0781 & 0.0771 &    41\\
PTF09hpq   &   00:41:12.91  &   -09:09:00.4  &    Gemini+GMOS   & 4000 & 3600 & 0.0529 & 0.0518 &    34\\
PTF09hql   &   01:55:29.30  &   +13:13:50.9  &    Gemini+GMOS   & 3000 & 3000 & 0.0499 & 0.0490 &    28\\
PTF09hqp   &   03:10:30.61  &   +35:05:02.2  &    Gemini+GMOS   & 3000 & 3000 & 0.0218 & 0.0212 &    37\\
PTF09ifh   &   04:10:35.27  &   +27:23:32.1  &    Gemini+GMOS   & 3000 & 3000 & 0.0788 & 0.0784 &    32\\
PTF10aaiw   &   01:09:21.02  &   +15:44:06.7  &    Gemini+GMOS   & 4000 & 3600 & 0.0600 & 0.0589 &    30\\
PTF10accd   &   02:13:30.50  &   +46:41:38.4  &    Gemini+GMOS   & 4000 & 4000 & 0.0348 & 0.0341 &    27\\
PTF10acnz   &   11:44:56.31  &   +58:39:46.1  &    Gemini+GMOS   & 3600 & 3600 & 0.0615 & 0.0620 &    21\\
PTF10bxs   &   10:58:44.28  &   +32:27:08.3  &    Gemini+GMOS   & 3600 & 3600 & 0.0727 & 0.0737 &    34\\
PTF10duz   &   12:51:40.03  &   +14:26:29.4  &    Gemini+GMOS   & 3600 & 3600 & 0.0640 & 0.0650 &    28\\
PTF10fps   &   13:29:24.26  &   +11:47:49.2  &    Gemini+GMOS   & 4000 & 3600 & 0.0214 & 0.0224 &    32\\
PTF10fxl   &   16:52:48.63  &   +51:03:41.0  &    Gemini+GMOS   & 3600 & 3600 & 0.0296 & 0.0295 &    41\\
PTF10gmd   &   11:57:18.96  &   +57:11:58.6  &    Gemini+GMOS   & 3600 & 3600 & 0.0552 & 0.0557 &    36\\
PTF10gmg   &   16:24:58.54  &   +51:02:22.2  &    Gemini+GMOS   & 3600 & 3600 & 0.0628 & 0.0628 &    28\\
PTF10hdn   &   14:52:24.58  &   +47:28:34.3  &    Gemini+GMOS   & 3600 & 3600 & 0.0699 & 0.0703 &    22\\
PTF10hdv   &   12:07:45.43  &   +41:29:28.7  &    Keck+LRIS   & 1800 & 1800 & 0.0527 & 0.0535 &    5\\
PTF10hjw   &   12:52:18.00  &   -07:01:06.7  &    Gemini+GMOS   & 3600 & 3600 & 0.0417 & 0.0428 &    30\\
PTF10hml   &   13:19:50.66  &   +41:58:57.4  &    Gemini+GMOS   & 3600 & 3600 & 0.0534 & 0.0541 &    15\\
PTF10icb   &   12:54:49.78  &   +58:52:56.6  &    Gemini+GMOS   & 2400 & 2400 & 0.0085 & 0.0089 &    31\\
PTF10iyc   &   17:09:22.03  &   +44:23:30.1  &    Gemini+GMOS   & 3600 & 3600 & 0.0594 & 0.0593 &    32\\
PTF10jdw   &   15:41:59.95  &   +47:35:29.8  &    Gemini+GMOS   & 3600 & 3600 & 0.0766 & 0.0768 &    32\\
PTF10jtp   &   17:10:58.51  &   +39:28:28.9  &    Gemini+GMOS   & 3600 & 3600 & 0.0670 & 0.0669 &    23\\
PTF10mwb   &   17:17:50.04  &   +40:52:52.7  &    Gemini+GMOS   & 4000 & 4000 & 0.0310 & 0.0309 &    11\\
PTF10nkd   &   15:45:21.41  &   +52:13:50.5  &    Gemini+GMOS   & 3600 & 3600 & 0.0676 & 0.0678 &    31\\
PTF10nlg   &   16:50:34.55  &   +60:16:34.6  &    Gemini+GMOS   & 3600 & 3600 & 0.0562 & 0.0561 &    27\\
PTF10pvi\tnote{c}   &   22:02:02.42  &   +14:32:07.1  &    WHT+ISIS   & 3600 & 3600 & 0.0802 & 0.0790 &    12\\
PTF10qjl   &   16:39:59.23  &   +12:06:26.6  &    WHT+ISIS   & 1800 & 1800 & 0.0576 & 0.0577 &    6\\
PTF10qjq   &   17:07:12.36  &   +35:30:35.6  &    Gemini+GMOS   & 3600 & 3600 & 0.0284 & 0.0283 &    24\\
PTF10qkf   &   23:14:23.04  &   +10:45:17.3  &    WHT+ISIS   & 3600 & 3600 & 0.0804 & 0.0792 &    19\\
PTF10qkv\tnote{c}   &   17:11:52.75  &   +27:22:20.3  &    WHT+ISIS   & 1200 & 1200 & 0.0611 & 0.0610 &    18\\
PTF10qky\tnote{c}   &   22:17:49.08  &   +05:25:23.5  &    Gemini+GMOS   & 3600 & 3600 & 0.0742 & 0.0730 &    38\\
PTF10qny\tnote{c}   &   16:09:30.67  &   +22:20:10.0  &    Gemini+GMOS   & 4000 & 4000 & 0.0333 & 0.0335 &    30\\
PTF10qsc   &   21:34:21.12  &   -05:03:46.1  &    WHT+ISIS   & 3600 & 3600 & 0.0879 & 0.0868 &    8\\
PTF10qwg   &   02:42:09.96  &   +02:26:51.0  &    Gemini+GMOS   & 4000 & 3600 & 0.0679 & 0.0671 &    37\\
PTF10rab\tnote{c}   &   01:47:07.46  &   -00:02:58.9  &    Keck+LRIS   & 1800 & 1800 & 0.0850 & 0.0840 &    5\\
PTF10rbp\tnote{c}   &   01:16:37.78  &   -01:49:28.2  &    Gemini+GMOS   & 4000 & 4000 & 0.0823 & 0.0813 &    38\\
PTF10tce   &   23:19:09.74  &   +09:11:45.6  &    WHT+ISIS   & 5400 & 5400 & 0.0410 & 0.0398 &    10\\
PTF10trp   &   21:28:08.01  &   +09:51:14.0  &    Gemini+GMOS   & 4000 & 4000 & 0.0489 & 0.0478 &    27\\
PTF10twd   &   23:00:14.23  &   +20:47:59.3  &    Gemini+GMOS   & 4000 & 4000 & 0.0734 & 0.0722 &    25\\
PTF10ubm   &   00:01:59.23  &   +21:49:31.1  &    Gemini+GMOS   & 3600 & 3600 & 0.0701 & 0.0689 &    27\\
PTF10viq   &   22:20:19.46  &   +17:03:25.2  &    WHT+ISIS   & 1800 & 1800 & 0.0342 & 0.0330 &    33\\
PTF10wnm   &   00:22:03.70  &   +27:02:21.5  &    WHT+ISIS   & 2400 & 2400 & 0.0663 & 0.0652 &    18\\
PTF10wnq   &   00:49:10.10  &   +32:08:14.3  &    WHT+ISIS   & 2400 & 2400 & 0.0691 & 0.0681 &    19\\
PTF10wof\tnote{c}   &   23:32:41.42  &   +15:21:31.7  &    WHT+ISIS   & 3000 & 3000 & 0.0530 & 0.0518 &    13\\
PTF10wor   &   22:32:41.67  &   -09:27:46.8  &    Gemini+GMOS   & 4000 & 3600 & 0.0568 & 0.0556 &    31\\
PTF10wos   &   21:44:39.36  &   -05:25:22.4  &    Gemini+GMOS   & 4000 & 4000 & 0.0820 & 0.0809 &    38\\
PTF10xyt   &   23:19:02.43  &   +13:47:26.2  &    Gemini+GMOS   & 3600 & 3600 & 0.0490 & 0.0478 &    37\\
PTF10yer   &   21:29:01.37  &   -01:25:51.6  &    Gemini+GMOS   & 4000 & 4000 & 0.0528 & 0.0517 &    53\\
PTF10ygu\tnote{c}   &   09:37:29.83  &   +23:09:41.8  &    Lick+KAST   & 600 & 600 & 0.0259 & 0.0269 &    14\\
PTF10yux   &   23:24:13.92  &   +07:13:39.7  &    Gemini+GMOS   & 3600 & 3600 & 0.0578 & 0.0566 &    38\\
PTF10zbk   &   02:29:34.46  &   +22:20:00.2  &    Gemini+GMOS   & 4000 & 4000 & 0.0642 & 0.0634 &    30\\
PTF10zdk   &   02:14:07.37  &   +23:37:49.1  &    WHT+ISIS   & 2400 & 2400 & 0.0322 & 0.0314 &    13\\
PTF10zgy   &   02:38:43.77  &   +14:06:11.9  &    Gemini+GMOS   & 4000 & 4000 & 0.0443 & 0.0435 &    25\\
PTF11apk   &   10:21:00.53  &   +21:42:59.8  &    Gemini+GMOS   & 3600 & 3600 & 0.0405 & 0.0416 &    28\\
PTF11atu   &   15:31:14.91  &   -00:46:45.8  &    Gemini+GMOS   & 4000 & 4000 & 0.0774 & 0.0779 &    23\\
PTF11bas   &   13:16:47.78  &   +43:31:13.4  &    Lick+KAST   & 2400 & 2400 & 0.0863 & 0.0870 &    15\\
PTF11bju   &   11:56:14.40  &   +25:21:14.8  &    Gemini+GMOS   & 4000 & 4000 & 0.0323 & 0.0333 &    31\\
PTF11htb   &   21:55:37.01  &   +00:41:29.4  &    WHT+ISIS   & 5400 & 5400 & 0.0493 & 0.0481 &    9\\
PTF11khk   &   17:02:58.03  &   +40:31:28.6  &    WHT+ISIS   & 3600 & 3600 & 0.0306 & 0.0305 &    26\\
PTF11kjn   &   22:45:03.93  &   +33:59:46.0  &    WHT+ISIS   & 3600 & 3600 & 0.0234 & 0.0223 &    29\\
PTF11kx   &   08:09:13.20  &   +46:18:42.8  &    Gemini+GMOS   & 4000 & 4000 & 0.0467 & 0.0472 &    39\\
PTF11lih   &   23:10:58.93  &   +31:51:36.0  &    WHT+ISIS   & 3600 & 3600 & 0.0720 & 0.0709 &    19\\
PTF11mty   &   21:34:05.21  &   +10:25:24.6  &    WHT+ISIS   & 3600 & 3600 & 0.0770 & 0.0759 &    20\\
PTF11okh   &   23:06:04.85  &   +34:06:27.7  &    WHT+ISIS   & 3600 & 3600 & 0.0191 & 0.0180 &    26\\
PTF11opu   &   22:07:50.02  &   +27:47:47.0  &    WHT+ISIS   & 3600 & 3600 & 0.0649 & 0.0638 &    9\\
PTF11pfm   &   21:44:13.80  &   +00:56:34.1  &    Keck+LRIS   & 1800 & 1800 & 0.0795 & 0.0784 &    8\\
PTF11v   &   15:08:23.42  &   +49:39:57.6  &    Gemini+GMOS   & 2400 & 2400 & 0.0373 & 0.0376 &    29\\
PTF11vl   &   16:28:40.27  &   +27:43:39.7  &    Gemini+GMOS   & 4000 & 4000 & 0.0454 & 0.0455 &    28\\
\hline
\label{obs-log}
\end{tabular}
\begin{tablenotes}
	\item[a] The exposure time for red or blue gratings/grisms in unit of second.
	\item[b] Here we demonstrate two redshifts based on heliocentric or CMB frame.
    \item[c] `Galaxy Zoo Supernovae project' discovered SNe.
\end{tablenotes}
\end{threeparttable}
\end{table*}

\begin{table*}
\centering
\caption{The host photometric properties in this paper.}
\begin{threeparttable}[b]
\renewcommand{\arraystretch}{0.91}
\scriptsize
\begin{tabular}{lrrrrrr}
\hline\hline
\multicolumn{7}{c}{Host photometric properties\tnote{a}}\\
\hline
SN name &
\multicolumn{1}{r}{\mstellar--} &
\multicolumn{1}{r}{\mstellar}	&
\multicolumn{1}{r}{\mstellar+} 	&
\multicolumn{1}{r}{logSFR--} 	&
\multicolumn{1}{r}{logSFR} 		&
\multicolumn{1}{r}{logSFR+}		\\
 &
										 		&
\multicolumn{1}{r}{($\mathrm{M_\odot}$)} 		&
										 		&
												&
\multicolumn{1}{r}{($\mathrm{M_\odot yr^{-1}}$)}&
												\\
\hline 
 PTF09dav & 10.20 & 10.28 & 10.46 &  0.47 &  0.58 &  0.65 \\
 PTF09dlc &  8.93 &  8.94 &  9.09 & -0.67 & -0.62 & -0.60 \\
 PTF09dnl &  7.98 &  8.06 &  8.39 & \nodata & \nodata & -0.76 \\
 PTF09dnp & 10.55 & 10.71 & 10.73 & \nodata & \nodata & \nodata \\
 PTF09dxo & 10.20 & 10.72 & 10.80 & \nodata & -0.38 & -0.36 \\
 PTF09dxw &  9.45 &  9.48 &  9.67 & -0.36 & -0.21 & -0.21 \\
 PTF09dza & 10.27 & 10.43 & 10.56 &  0.19 &  0.50 &  0.60 \\
 PTF09edr & 11.27 & 11.30 & 11.41 & \nodata & \nodata &  0.18 \\
 PTF09eoi &  9.24 &  9.39 &  9.47 & \nodata & \nodata & \nodata \\
 PTF09fox & 10.27 & 10.36 & 10.43 &  0.43 &  0.47 &  0.56 \\
 PTF09foz & 10.40 & 10.56 & 10.56 & \nodata & \nodata & \nodata \\
 PTF09gce & 10.82 & 10.86 & 10.95 & \nodata & -0.28 &  0.40 \\
 PTF09gon & 11.22 & 11.23 & 11.34 & \nodata & \nodata & \nodata \\
 PTF09gul & 10.99 & 11.11 & 11.14 & \nodata & -0.32 & -0.04 \\
 PTF09hpl & 10.62 & 10.69 & 10.87 &  0.91 &  1.03 &  1.08 \\
 PTF09hpq & 10.38 & 10.38 & 10.42 & \nodata & \nodata & \nodata \\
 PTF09hql &  9.40 &  9.50 &  9.61 & -0.32 & -0.25 & -0.15 \\
 PTF09hqp & \nodata & 10.90 & \nodata & \nodata &  0.96 & \nodata \\
 PTF09ifh & \nodata & 10.80 & \nodata & \nodata &  0.58 & \nodata \\
PTF10aaiw & 10.64 & 10.66 & 10.76 & \nodata & -0.44 &  0.04 \\
PTF10accd &  9.05 &  9.13 &  9.51 & \nodata & \nodata & -0.28 \\
PTF10acnz & 10.63 & 10.64 & 10.67 & \nodata & \nodata & \nodata \\
 PTF10bxs & 10.62 & 10.72 & 10.91 &  0.39 &  0.90 &  0.95 \\
 PTF10duz & 10.15 & 10.18 & 10.25 &  0.56 &  0.61 &  0.62 \\
 PTF10fps & 10.39 & 10.49 & 10.51 & \nodata & \nodata & \nodata \\
 PTF10fxl & 10.53 & 11.08 & 11.17 & \nodata & -0.02 &  0.49 \\
 PTF10gmd & 10.30 & 10.36 & 10.38 & \nodata & \nodata & \nodata \\
 PTF10gmg &  9.67 &  9.77 &  9.80 &  0.05 &  0.07 &  0.13 \\
 PTF10hdn &  8.65 &  8.79 &  8.95 & -0.28 & -0.13 &  0.05 \\
 PTF10hdv &  7.51 &  7.72 &  7.97 & -1.43 & -1.21 & -0.94 \\
 PTF10hjw & \nodata & 10.00 & \nodata & \nodata &  0.43 & \nodata \\
 PTF10hml &  9.83 &  9.95 & 10.38 & \nodata & \nodata &  1.06 \\
 PTF10icb &  9.54 &  9.60 &  9.75 & -0.15 & -0.05 & -0.02 \\
 PTF10iyc & 10.52 & 10.56 & 11.20 & \nodata & \nodata &  1.25 \\
 PTF10jdw & 10.81 & 10.90 & 11.11 &  0.51 &  1.12 &  1.17 \\
 PTF10jtp & 11.13 & 11.14 & 11.24 & -0.06 &  0.09 &  0.76 \\
 PTF10mwb &  8.98 &  9.31 &  9.55 & \nodata & -0.37 & -0.29 \\
 PTF10nkd &  9.63 &  9.70 &  9.85 & -0.04 &  0.03 &  0.09 \\
 PTF10nlg & \nodata & 10.01 & \nodata & \nodata &  0.45 & \nodata \\
 PTF10pvi & 10.17 & 10.31 & 10.45 &  0.48 &  0.54 &  0.65 \\
 PTF10qjl &  8.00 &  8.46 &  8.67 & -0.96 & -0.48 & -0.25 \\
 PTF10qjq &  9.49 &  9.56 &  9.62 &  0.57 &  0.65 &  0.71 \\
 PTF10qkf & 10.37 & 10.40 & 10.50 &  0.50 &  0.58 &  0.63 \\
 PTF10qkv & 10.43 & 10.50 & 10.61 & -0.11 &  0.59 &  0.66 \\
 PTF10qky & 10.46 & 10.53 & 10.69 &  0.75 &  0.86 &  0.92 \\
 PTF10qny & 10.17 & 10.69 & 10.81 & \nodata & -0.40 &  0.48 \\
 PTF10qsc &  9.77 &  9.83 &  9.90 &  0.17 &  0.18 &  0.25 \\
 PTF10qwg &  9.91 & 10.00 & 10.22 & -0.19 &  0.23 &  0.27 \\
 PTF10rab &  7.14 &  8.15 &  8.87 & \nodata & \nodata & -0.63 \\
 PTF10rbp & 10.70 & 11.16 & 11.25 & \nodata &  0.73 &  1.29 \\
 PTF10tce & 10.38 & 10.51 & 10.69 &  0.62 &  0.73 &  0.81 \\
 PTF10trp & 10.14 & 10.27 & 10.40 &  0.41 &  0.47 &  0.58 \\
 PTF10twd &  9.86 & 10.00 & 10.05 &  0.21 &  0.23 &  0.31 \\
 PTF10ubm & 10.00 & 10.11 & 10.17 &  0.37 &  0.39 &  0.48 \\
 PTF10viq & 10.81 & 10.94 & 10.99 &  0.18 &  0.26 &  1.04 \\
 PTF10wnm & 10.52 & 10.56 & 10.59 &  0.70 &  0.71 &  0.75 \\
 PTF10wnq & 11.29 & 11.36 & 11.42 & \nodata & \nodata & \nodata \\
 PTF10wof &  9.93 & 10.09 & 10.19 &  0.19 &  0.25 &  0.37 \\
 PTF10wor & 11.45 & 11.51 & 11.57 & \nodata & \nodata & \nodata \\
 PTF10wos & 11.23 & 11.25 & 11.30 &  0.07 &  0.16 &  0.20 \\
 PTF10xyt &  9.49 &  9.69 &  9.79 & -0.25 & -0.24 & -0.06 \\
 PTF10yer & 10.58 & 10.66 & 10.87 &  0.76 &  0.91 &  0.95 \\
 PTF10ygu & 10.70 & 11.40 & 11.44 & \nodata & \nodata &  0.14 \\
 PTF10yux & 11.23 & 11.29 & 11.33 & \nodata & \nodata & \nodata \\
 PTF10zbk & 10.71 & 10.80 & 10.85 & -0.64 &  0.17 &  0.80 \\
 PTF10zdk &  9.08 &  9.15 &  9.64 &  0.02 &  0.23 &  0.29 \\
 PTF10zgy & \nodata & 11.39 & \nodata & \nodata &  1.38 & \nodata \\
 PTF11apk & 10.74 & 10.90 & 10.91 & \nodata & \nodata & \nodata \\
 PTF11atu &  9.39 &  9.77 &  9.99 & \nodata & -0.13 & -0.01 \\
 PTF11bas &  9.71 &  9.75 &  9.80 &  0.12 &  0.16 &  0.18 \\
 PTF11bju &  9.51 &  9.58 &  9.61 & -0.16 & -0.14 & -0.08 \\
 PTF11htb &  8.94 &  8.99 &  9.02 & -0.72 & -0.71 & -0.66 \\
 PTF11khk & 10.53 & 10.70 & 10.74 & \nodata & \nodata & \nodata \\
 PTF11kjn & 11.56 & 11.62 & 11.67 & \nodata & \nodata & \nodata \\
  PTF11kx & 10.20 & 10.26 & 10.48 &  0.42 &  0.59 &  0.60 \\
 PTF11lih & 10.64 & 10.70 & 10.86 &  0.85 &  0.96 &  0.99 \\
 PTF11mty &  9.86 &  9.93 & 10.08 &  0.24 &  0.32 &  0.34 \\
 PTF11okh & 11.35 & 11.40 & 11.45 & \nodata & \nodata &  0.03 \\
 PTF11opu &  9.74 &  9.82 &  9.97 &  0.05 &  0.12 &  0.21 \\
 PTF11pfm &  8.98 &  9.12 &  9.20 & -0.68 & -0.66 & -0.57 \\
   PTF11v & 11.05 & 11.06 & 11.23 & \nodata & \nodata & \nodata \\
  PTF11vl &  9.64 & 10.07 & 10.30 & \nodata &  0.29 &  0.42 \\
  \hline
\label{host_para_phot}
\end{tabular}
\begin{tablenotes}
	\item[a] The host parameters determined photometrically by \textsc{z-peg}.
\end{tablenotes}
\end{threeparttable}
\end{table*}

\begin{table*}
\centering
\caption{The host spectroscopic properties in this paper.}
\begin{threeparttable}[b]
\renewcommand{\arraystretch}{0.91}
\scriptsize
\begin{tabular}{lrrrrc}
\hline\hline
\multicolumn{6}{c}{Host spectroscopic properties\tnote{a}} \\
\hline
SN name 						&
\multicolumn{1}{c}{12+log(O/H)} &
\multicolumn{1}{c}{logSFR} 		&
\multicolumn{1}{c}{[M/H]} 		&
\multicolumn{1}{c}{logAge} 		&
AGN\tnote{b} 					\\
 												&
					 							&
\multicolumn{1}{c}{($\mathrm{M_\odot yr^{-1}}$)}& 
 												&
(Gyr) 											&
 												\\
\hline 
 PTF09dav & $8.698 \pm 0.073$ & $ 0.0778 \pm  0.2001$ & -0.476 &  0.406 & 0 \\
 PTF09dlc & $8.182 \pm 0.069$ & $-0.7730 \pm  0.2002$ & -1.106 &  0.832 & 0 \\
 PTF09dnl & \nodata & $-2.0440 \pm  0.2003$ & -1.107 &  0.808 & 0 \\
 PTF09dnp & $8.758 \pm 0.081$ & $<-1.5310$ & -0.130 &  1.110 & 1 \\
 PTF09dxo & $8.732 \pm 0.075$ & $-0.1270 \pm  0.2001$ &  0.033 &  0.799 & 0 \\
 PTF09dxw & $8.627 \pm 0.072$ & $-0.9420 \pm  0.2001$ & -0.598 &  0.742 & 0 \\
 PTF09dza & $8.688 \pm 0.072$ & $ 0.1926 \pm  0.2001$ & -0.197 &  0.514 & 0 \\
 PTF09edr & $8.760 \pm 0.083$ & $-0.5293 \pm  0.2002$ & -0.111 &  1.045 & 1 \\
 PTF09eoi & $8.518 \pm 0.096$ & $-0.8333 \pm  0.2008$ & -0.454 &  0.211 & 0 \\
 PTF09fox & $8.664 \pm 0.071$ & $ 0.5646 \pm  0.2000$ & -0.216 &  0.243 & 0 \\
 PTF09foz & \nodata & $<-1.7998$ & -0.033 &  0.735 & 0 \\
 PTF09gce & $8.835 \pm 0.073$ & $-0.2271 \pm  0.2001$ & -0.189 &  0.745 & 0 \\
 PTF09gon & \nodata & $<-1.4682$ &  0.052 &  0.994 & 0 \\
 PTF09gul & $8.804 \pm 0.074$ & $-0.4125 \pm  0.2001$ & -0.212 &  0.831 & 1 \\
 PTF09hpl & $8.710 \pm 0.072$ & $ 0.5366 \pm  0.2001$ & -0.292 &  0.674 & 0 \\
 PTF09hpq & \nodata & $<-2.1847$ &  0.122 &  0.932 & 0 \\
 PTF09hql & $8.573 \pm 0.081$ & $-1.1066 \pm  0.2001$ & -1.132 &  0.564 & 0 \\
 PTF09hqp & $8.826 \pm 0.070$ & $ 0.8022 \pm  0.2000$ & -0.132 &  0.585 & 0 \\
 PTF09ifh & $8.724 \pm 0.093$ & $ 0.1958 \pm  0.2008$ & -0.513 &  0.837 & 0 \\
PTF10aaiw & $8.854 \pm 0.070$ & $-0.6567 \pm  0.2001$ & -0.349 &  0.944 & 0 \\
PTF10accd & $8.362 \pm 0.087$ & $-0.8841 \pm  0.2002$ & -1.046 &  0.374 & 0 \\
PTF10acnz & $8.799 \pm 0.079$ & $-0.7896 \pm  0.2023$ &  0.074 &  0.890 & 1 \\
 PTF10bxs & $8.718 \pm 0.075$ & $ 0.4785 \pm  0.2002$ & -0.530 &  0.771 & 0 \\
 PTF10duz & $8.571 \pm 0.074$ & $ 0.3213 \pm  0.2001$ & -0.511 &  0.398 & 0 \\
 PTF10fps & $8.751 \pm 0.071$ & $ 0.4652 \pm  0.2000$ &  0.083 &  0.413 & 0 \\
 PTF10fxl & $8.752 \pm 0.070$ & $ 0.4853 \pm  0.2000$ & -0.162 &  0.578 & 0 \\
 PTF10gmd & $8.748 \pm 0.084$ & $<-1.9430$ & -0.075 &  0.722 & 1 \\
 PTF10gmg & $8.396 \pm 0.078$ & $-0.3354 \pm  0.2001$ & -1.148 &  0.994 & 0 \\
 PTF10hdn & $8.224 \pm 0.090$ & $-0.2529 \pm  0.2002$ & -1.297 &  0.695 & 0 \\
 PTF10hdv & $8.101 \pm 0.079$ & $-1.1635 \pm  0.2005$ & -1.099 &  0.734 & 0 \\
 PTF10hjw & $8.618 \pm 0.073$ & $-0.0162 \pm  0.2001$ & -1.081 &  0.654 & 0 \\
 PTF10hml & $8.506 \pm 0.077$ & $ 0.0052 \pm  0.2001$ & -0.210 &  0.249 & 0 \\
 PTF10icb & $8.438 \pm 0.074$ & $-0.1586 \pm  0.2001$ & -0.474 &  0.348 & 0 \\
 PTF10iyc & $8.740 \pm 0.087$ & $-0.4100 \pm  0.2002$ & -0.306 &  1.052 & 0 \\
 PTF10jdw & $8.738 \pm 0.089$ & $ 0.4382 \pm  0.2006$ & -0.605 &  0.765 & 0 \\
 PTF10jtp & $8.779 \pm 0.078$ & $ 0.9699 \pm  0.2001$ & -0.201 &  0.827 & 0 \\
 PTF10mwb & \nodata & $-1.3801 \pm  0.2001$ & -1.179 &  0.605 & 0 \\
 PTF10nkd & $8.570 \pm 0.082$ & $-0.6267 \pm  0.2003$ & -1.116 &  0.614 & 0 \\
 PTF10nlg & $8.552 \pm 0.071$ & $-0.2332 \pm  0.2000$ &  0.013 & -0.252 & 0 \\
 PTF10pvi & $8.621 \pm 0.167$ & $ 0.1395 \pm  0.2034$ & -0.835 &  0.826 & 0 \\
 PTF10qjl & $8.191 \pm 0.214$ & $ 0.0953 \pm  0.2022$ & -0.905 &  0.730 & 0 \\
 PTF10qjq & $8.678 \pm 0.070$ & $ 0.5050 \pm  0.2000$ & -1.005 &  0.621 & 0 \\
 PTF10qkf & $8.680 \pm 0.087$ & $-0.0993 \pm  0.2003$ & -0.876 &  0.723 & 0 \\
 PTF10qkv & $8.755 \pm 0.082$ & $ 0.0145 \pm  0.2003$ & -0.209 &  0.810 & 0 \\
 PTF10qky & $8.675 \pm 0.071$ & $ 0.7003 \pm  0.2000$ & -0.073 &  0.312 & 0 \\
 PTF10qny & $8.645 \pm 0.078$ & $ 0.7040 \pm  0.2002$ & -0.408 &  0.404 & 0 \\
 PTF10qsc & \nodata & $-0.3717 \pm  0.2006$ & -0.320 &  0.755 & 0 \\
 PTF10qwg & $8.727 \pm 0.072$ & $-0.0504 \pm  0.2000$ & -0.363 &  0.905 & 0 \\
 PTF10rab & $8.098 \pm 0.067$ & $-1.3371 \pm  0.2000$ & \nodata & \nodata & 0 \\
 PTF10rbp & $8.816 \pm 0.075$ & $ 0.2401 \pm  0.2001$ & -0.153 &  0.665 & 0 \\
 PTF10tce & $8.663 \pm 0.145$ & $-0.1912 \pm  0.2030$ & -0.185 &  0.707 & 0 \\
 PTF10trp & $8.623 \pm 0.072$ & $-0.0145 \pm  0.2001$ & -0.860 &  0.534 & 0 \\
 PTF10twd & $8.505 \pm 0.072$ & $ 0.0067 \pm  0.2000$ & -0.693 &  0.572 & 0 \\
 PTF10ubm & $8.664 \pm 0.070$ & $ 0.2402 \pm  0.2000$ & -0.573 &  0.626 & 0 \\
 PTF10viq & $8.822 \pm 0.076$ & $ 1.0179 \pm  0.2003$ & -0.025 &  0.558 & 0 \\
 PTF10wnm & $8.697 \pm 0.119$ & $ 0.4628 \pm  0.2018$ & -0.131 &  0.482 & 0 \\
 PTF10wnq & \nodata & $<-1.0496$ & -0.130 &  1.035 & 0 \\
 PTF10wof & $8.575 \pm 0.100$ & $-0.1685 \pm  0.2011$ & -1.035 &  0.803 & 0 \\
 PTF10wor & $8.903 \pm 0.057$ & $-0.7775 \pm  0.2008$ &  0.071 &  0.939 & 1 \\
 PTF10wos & \nodata & $<-2.0219$ & -0.117 &  0.916 & 0 \\
 PTF10xyt & $8.397 \pm 0.078$ & $-0.4585 \pm  0.2001$ & -1.047 &  0.443 & 0 \\
 PTF10yer & $8.624 \pm 0.071$ & $ 0.4392 \pm  0.2000$ & -1.011 &  0.781 & 0 \\
 PTF10ygu & \nodata & $ 0.4154 \pm  0.2041$ & -0.420 &  0.834 & 0 \\
 PTF10yux & $8.711 \pm 0.085$ & $-0.6406 \pm  0.2002$ & -0.199 &  0.922 & 1 \\
 PTF10zbk & $8.768 \pm 0.071$ & $-0.0288 \pm  0.2000$ & -0.203 &  0.820 & 0 \\
 PTF10zdk & $8.431 \pm 0.101$ & $-0.4892 \pm  0.2001$ & -0.402 &  0.005 & 0 \\
 PTF10zgy & $8.827 \pm 0.072$ & $ 0.9997 \pm  0.2000$ &  0.010 &  0.601 & 1 \\
 PTF11apk & $8.829 \pm 0.074$ & $-0.7912 \pm  0.2002$ & -0.005 &  0.981 & 1 \\
 PTF11atu & $8.559 \pm 0.086$ & $-0.6486 \pm  0.2005$ & -0.842 &  0.711 & 0 \\
 PTF11bas & \nodata & $ 0.0040 \pm  0.2270$ & -1.084 &  0.735 & 0 \\
 PTF11bju & $8.363 \pm 0.072$ & $-0.4429 \pm  0.2000$ & -0.893 &  0.304 & 0 \\
 PTF11htb & $8.235 \pm 0.181$ & $-0.6434 \pm  0.2006$ & -1.135 &  0.640 & 0 \\
 PTF11khk & $8.777 \pm 0.081$ & $ 0.4498 \pm  0.2012$ &  0.095 &  0.599 & 0 \\
 PTF11kjn & $8.816 \pm 0.074$ & $-0.3221 \pm  0.2010$ &  0.108 &  0.993 & 1 \\
  PTF11kx & $8.679 \pm 0.072$ & $-0.1617 \pm  0.2001$ & -0.737 &  0.303 & 0 \\
 PTF11lih & $8.747 \pm 0.150$ & $ 1.4264 \pm  0.2038$ & -0.212 &  0.907 & 0 \\
 PTF11mty & $8.584 \pm 0.082$ & $ 0.7431 \pm  0.2004$ & -0.100 &  0.346 & 0 \\
 PTF11okh & $8.811 \pm 0.074$ & $-0.0839 \pm  0.2001$ &  0.117 &  1.056 & 0 \\
 PTF11opu & $8.319 \pm 0.175$ & $ 0.1381 \pm  0.2024$ & -0.857 &  0.721 & 0 \\
 PTF11pfm & $8.406 \pm 0.070$ & $-1.1024 \pm  0.2000$ & -0.193 &  0.261 & 0 \\
   PTF11v & $8.795 \pm 0.076$ & $-0.7488 \pm  0.2004$ & -0.014 &  1.078 & 1 \\
  PTF11vl & $8.581 \pm 0.093$ & $-0.6123 \pm  0.2004$ & -1.020 &  0.664 & 0 \\
  \hline
\label{host_para_spec}
\end{tabular}
\begin{tablenotes}
	\item[a] The host parameters determined either from emission line or stellar continuum measurements.
	\item[b] The AGN tags for the host galaxies in this work. Normal galaxies are labeled as 0 and
	AGN hosts are labeled as 1.
\end{tablenotes}
\end{threeparttable}
\end{table*}

\begin{table*}
\centering
\caption{The SN photometric properties in this paper.}
\begin{threeparttable}[b]
\renewcommand{\arraystretch}{0.91}
\scriptsize
\begin{tabular}{llrrr}
\hline\hline
\multicolumn{5}{c}{SN photometric properties\tnote{a}}\\
\hline
SN name &
\multicolumn{1}{c}{LC source\tnote{b}} &
\multicolumn{1}{c}{stretch ($s$)} &
\multicolumn{1}{c}{colour (\col)} &
\multicolumn{1}{c}{$m_B$ (mag)}  \\
\hline
 PTF09dav &P48& $0.64 \pm 0.04$ & \nodata & \nodata \\
 PTF09dlc &P48; FT& $1.06 \pm 0.01$ & $ 0.02 \pm  0.01$ & $18.17 \pm  0.02$ \\
 PTF09dnl &P48; LT& $1.05 \pm 0.02$ & $-0.02 \pm  0.01$ & $15.85 \pm  0.02$ \\
 PTF09dnp &P48; FT& $0.98 \pm 0.03$ & $ 0.11 \pm  0.02$ & $17.09 \pm  0.02$ \\
 PTF09dxo &P48& $1.08 \pm 0.04$ & \nodata & \nodata \\
 PTF09dxw &P48& $0.72 \pm 0.05$ & \nodata & \nodata \\
 PTF09dza &P48& \nodata & \nodata & \nodata \\
 PTF09edr &P48& $0.81 \pm 0.11$ & \nodata & \nodata \\
 PTF09eoi &P48& \nodata & \nodata & \nodata \\
 PTF09fox &P48; LT& $0.95 \pm 0.04$ & $ 0.03 \pm  0.04$ & $18.34 \pm  0.06$ \\
 PTF09foz &P48; LT& $0.85 \pm 0.04$ & $ 0.00 \pm  0.06$ & $17.78 \pm  0.09$ \\
 PTF09gce &P48; P60& $1.06 \pm 0.05$ & $-0.03 \pm  0.03$ & $17.70 \pm  0.07$ \\
 PTF09gon &P48& \nodata & \nodata & \nodata \\
 PTF09gul &P48& $0.89 \pm 0.06$ & \nodata & \nodata \\
 PTF09hpl &P48& $0.89 \pm 0.05$ & \nodata & \nodata \\
 PTF09hpq &P48& $0.82 \pm 0.08$ & \nodata & \nodata \\
 PTF09hql &P48& $1.11 \pm 0.04$ & \nodata & \nodata \\
 PTF09hqp &P48& $1.16 \pm 0.03$ & \nodata & \nodata \\
 PTF09ifh &P48& \nodata & \nodata & \nodata \\
PTF10aaiw &\nodata& \nodata & \nodata & \nodata \\
PTF10accd &P48; LT& $1.03 \pm 0.02$ & $-0.11 \pm  0.02$ & $16.60 \pm  0.05$ \\
PTF10acnz &P48; FT& $1.10 \pm 0.01$ & $ 0.10 \pm  0.01$ & $17.98 \pm  0.01$ \\
 PTF10bxs &P48; P60& $1.08 \pm 0.05$ & $ 0.11 \pm  0.09$ & $18.81 \pm  0.12$ \\
 PTF10duz &P48; P60& $0.97 \pm 0.03$ & $-0.01 \pm  0.02$ & $18.06 \pm  0.03$ \\
 PTF10fps &P48; LT& $0.72 \pm 0.02$ & $ 0.12 \pm  0.03$ & $16.66 \pm  0.05$ \\
 PTF10fxl &P48; FT& $0.97 \pm 0.02$ & $ 0.10 \pm  0.01$ & $16.76 \pm  0.02$ \\
 PTF10gmd &P48; FT& $0.78 \pm 0.01$ & $ 0.09 \pm  0.01$ & $18.35 \pm  0.02$ \\
 PTF10gmg &P48; P60& $1.11 \pm 0.02$ & $-0.03 \pm  0.02$ & $17.91 \pm  0.03$ \\
 PTF10hdn &P48; LT& $1.09 \pm 0.03$ & \nodata & \nodata \\
 PTF10hdv &P48; FT& $1.09 \pm 0.01$ & $-0.03 \pm  0.01$ & $17.56 \pm  0.01$ \\
 PTF10hjw &P48; LT& $0.99 \pm 0.02$ & $-0.09 \pm  0.04$ & $16.99 \pm  0.06$ \\
 PTF10hml &P48; LT& $1.02 \pm 0.01$ & $-0.05 \pm  0.03$ & $17.71 \pm  0.03$ \\
 PTF10icb &P48; LT& $0.98 \pm 0.03$ & $ 0.06 \pm  0.02$ & $14.49 \pm  0.03$ \\
 PTF10iyc &P48; LT& $1.06 \pm 0.02$ & $-0.03 \pm  0.01$ & $17.74 \pm  0.01$ \\
 PTF10jdw &P48; LT& $0.84 \pm 0.02$ & \nodata & \nodata \\
 PTF10jtp &P48; LT& $0.87 \pm 0.01$ & $ 0.11 \pm  0.02$ & $18.69 \pm  0.03$ \\
 PTF10mwb &P48; LT& $0.92 \pm 0.02$ & $ 0.03 \pm  0.03$ & $16.82 \pm  0.04$ \\
 PTF10nkd &P48; LT& $0.92 \pm 0.03$ & \nodata & \nodata \\
 PTF10nlg &P48; LT& $0.98 \pm 0.03$ & $ 0.10 \pm  0.03$ & $18.59 \pm  0.03$ \\
 PTF10pvi &P48; LT& $1.03 \pm 0.02$ & $-0.08 \pm  0.02$ & $18.43 \pm  0.03$ \\
 PTF10qjl &P48; LT& $0.96 \pm 0.02$ & $-0.11 \pm  0.02$ & $17.82 \pm  0.02$ \\
 PTF10qjq &P48; LT& $0.93 \pm 0.02$ & $ 0.00 \pm  0.02$ & $16.42 \pm  0.03$ \\
 PTF10qkf &P48; LT& $1.03 \pm 0.02$ & $ 0.10 \pm  0.03$ & $18.95 \pm  0.03$ \\
 PTF10qkv &P48; LT& $1.07 \pm 0.04$ & $ 0.17 \pm  0.03$ & $19.15 \pm  0.03$ \\
 PTF10qky &P48; LT& $1.07 \pm 0.02$ & $-0.06 \pm  0.02$ & $18.14 \pm  0.04$ \\
 PTF10qny &P48; P60& $1.10 \pm 0.07$ & $-0.02 \pm  0.02$ & $16.37 \pm  0.05$ \\
 PTF10qsc &P48; LT& $1.15 \pm 0.02$ & $-0.07 \pm  0.02$ & $18.54 \pm  0.02$ \\
 PTF10qwg &P48; LT& $0.93 \pm 0.04$ & $ 0.16 \pm  0.07$ & $18.58 \pm  0.10$ \\
 PTF10rab &P48; LT& $1.06 \pm 0.04$ & \nodata & \nodata \\
 PTF10rbp &P48; LT& $1.08 \pm 0.03$ & $ 0.09 \pm  0.02$ & $18.76 \pm  0.03$ \\
 PTF10tce &P48; LT& $1.08 \pm 0.02$ & $ 0.05 \pm  0.02$ & $17.19 \pm  0.03$ \\
 PTF10trp &P48; LT& $1.12 \pm 0.02$ & $ 0.59 \pm  0.02$ & $19.15 \pm  0.03$ \\
 PTF10twd &P48; LT& $1.08 \pm 0.02$ & $-0.07 \pm  0.01$ & $18.08 \pm  0.03$ \\
 PTF10ubm &P48; LT& $1.06 \pm 0.01$ & $ 0.01 \pm  0.01$ & $17.98 \pm  0.02$ \\
 PTF10viq &P48; LT& $1.13 \pm 0.01$ & $ 0.05 \pm  0.02$ & $16.62 \pm  0.03$ \\
 PTF10wnm &P48; LT& $1.04 \pm 0.02$ & $ 0.03 \pm  0.01$ & $18.20 \pm  0.02$ \\
 PTF10wnq &P48; LT& $0.93 \pm 0.02$ & $-0.16 \pm  0.02$ & $18.10 \pm  0.03$ \\
 PTF10wof &P48; LT& $0.99 \pm 0.03$ & $ 0.09 \pm  0.04$ & $17.95 \pm  0.07$ \\
 PTF10wor &P48; LT& $0.68 \pm 0.03$ & $ 0.55 \pm  0.05$ & $20.03 \pm  0.11$ \\
 PTF10wos &P48; LT& $0.93 \pm 0.05$ & $-0.04 \pm  0.02$ & $18.73 \pm  0.04$ \\
 PTF10xyt &P48; LT& $1.08 \pm 0.03$ & $ 0.22 \pm  0.03$ & $18.44 \pm  0.04$ \\
 PTF10yer &P48; P60& \nodata & \nodata & \nodata \\
 PTF10ygu &P48; LT& $1.01 \pm 0.02$ & $ 0.44 \pm  0.03$ & $17.27 \pm  0.04$ \\
 PTF10yux &P48; LT& $0.84 \pm 0.01$ & $ 0.17 \pm  0.02$ & $18.52 \pm  0.04$ \\
 PTF10zbk &P48; LT& $0.78 \pm 0.03$ & $ 0.11 \pm  0.06$ & $19.28 \pm  0.11$ \\
 PTF10zdk &P48; LT& $1.14 \pm 0.01$ & $ 0.06 \pm  0.02$ & $16.96 \pm  0.04$ \\
 PTF10zgy &P48; LT& $1.16 \pm 0.04$ & $ 0.15 \pm  0.03$ & $18.00 \pm  0.06$ \\
 PTF11apk &P48; LT& $0.75 \pm 0.04$ & $ 0.21 \pm  0.04$ & $18.32 \pm  0.05$ \\
 PTF11atu &P48; LT& $0.81 \pm 0.04$ & $-0.06 \pm  0.04$ & $18.62 \pm  0.10$ \\
 PTF11bas &P48; LT& $0.96 \pm 0.02$ & $ 0.06 \pm  0.01$ & $18.92 \pm  0.02$ \\
 PTF11bju &P48; LT& $1.15 \pm 0.01$ & $ 0.02 \pm  0.01$ & $16.62 \pm  0.01$ \\
 PTF11htb &P48; LT& $1.06 \pm 0.02$ & $ 0.01 \pm  0.02$ & $17.37 \pm  0.04$ \\
 PTF11khk &P48; LT& $0.59 \pm 0.02$ & $ 0.41 \pm  0.04$ & $17.97 \pm  0.06$ \\
 PTF11kjn &P48; LT& $0.62 \pm 0.01$ & $ 0.28 \pm  0.02$ & $17.31 \pm  0.03$ \\
  PTF11kx &P48& $1.04 \pm 0.02$ & \nodata & \nodata \\
 PTF11lih &P48; LT& $0.89 \pm 0.01$ & $ 0.10 \pm  0.02$ & $18.97 \pm  0.03$ \\
 PTF11mty &P48; LT& $1.05 \pm 0.03$ & $ 0.03 \pm  0.03$ & $18.44 \pm  0.04$ \\
 PTF11okh &P48; LT& $0.53 \pm 0.01$ & $ 0.56 \pm  0.03$ & $18.17 \pm  0.06$ \\
 PTF11opu &P48; LT& $1.34 \pm 0.03$ & $ 0.02 \pm  0.03$ & $19.39 \pm  0.05$ \\
 PTF11pfm &P48;   & \nodata & \nodata & \nodata \\
   PTF11v &P48; LT& $0.78 \pm 0.01$ & $-0.00 \pm  0.03$ & $17.05 \pm  0.04$ \\
  PTF11vl &P48; LT& \nodata & \nodata & \nodata \\
\hline
\label{sn_para}
\end{tabular}
\begin{tablenotes}
	\item[a] The SN properties derived by SiFTO light curve fitting code.
	\item[b] See Section~\ref{sec:sample-selection} for more information. 
\end{tablenotes}
\end{threeparttable}
\end{table*}

\label{lastpage}

\end{document}